%% file: main.tex
\documentclass[sigplan,nonacm]{acmart}

\setcopyright{none}
\renewcommand\footnotetextcopyrightpermission[1]{}

\usepackage{booktabs}
\usepackage{microtype}
\usepackage{subcaption}
\usepackage{xspace}
\usepackage{enumitem}
\usepackage{balance}
\usepackage{algorithm}
\usepackage{algpseudocode}
\usepackage{pifont}
\newcommand{\stepnum}[1]{\textcolor[HTML]{B2182B}{\ding{\numexpr171+#1\relax}}}
\usepackage{stfloats}

\newcommand{\sys}{CoWalk\xspace}

\newcommand{\pipea}{PipeANN\xspace}
\newcommand{\diska}{DiskANN\xspace}

\begin{document}

\title[Cooperative Walks for Agentic Batch Queries]{CoWalk: Cooperative Walks for\\
Agentic Batch Queries on Disk-Resident Graphs}

\author{Qiao Liang$^{1,2}$, Zhiguang Chen$^{1}$, Yue Yu$^{2}$, Nong Xiao$^{1}$}
\affiliation{%
  \institution{$^{1}$Sun Yat-sen University\qquad $^{2}$Pengcheng Laboratory}
  \country{}
}
\email{liangqao@mail2.sysu.edu.cn}

\begin{abstract}
Workloads such as query rewriting, deep research and group-relative policy optimization (GRPO) issue batches of semantically related vector queries.
Existing disk-resident approximate nearest neighbor (ANN) systems process these queries independently.
On three workloads, 31 to 67 percent of the I/Os of a batch fetch a page that another query of the batch has already fetched.
A page cache shared by the batch removes these I/Os, yet it makes a single-thread search only 1.1 to $1.2{\times}$ faster, because the search time goes to distance computation and the cache does not change the walk of any query.
We present \sys, which runs the queries of a batch as cooperating walks on one thread.
Every query keeps its own candidate list, a page is read once per batch, and the full-precision vector on every fetched page is ranked by every query of the batch.
A query therefore reaches the target recall with a shorter candidate list, which computes fewer distances and issues fewer I/Os.
At recall@10 of 0.95, \sys\ answers a batch on one thread 1.6 to $1.9{\times}$ faster than the fastest per-query baseline and issues 2.8 to $6.9{\times}$ fewer I/Os.
On a server it serves 3.0 to $6.9{\times}$ the throughput of that baseline at a similar batch latency.
\end{abstract}

\maketitle
\pagestyle{plain}

\begin{figure*}[!b]
  \centering
  \includegraphics[width=\textwidth]{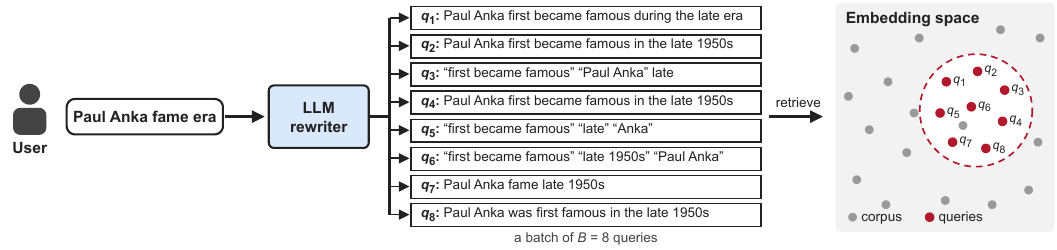}
  \caption{An LLM rewriter expands one user intent into a batch of
  paraphrased queries that cluster within a small region of embedding
  space (OpenResearcher trajectory 38476, $B{=}8$).}
  \label{fig:teaser}
\end{figure*}

\input{sections/01_introduction}
\input{sections/02_background_motivation}

\input{sections/03_design}
\input{sections/04_implementation}
\input{sections/05_evaluation}
\input{sections/06_related_work}
\input{sections/07_discussion}
\input{sections/08_conclusion}

\bibliographystyle{ACM-Reference-Format}
\bibliography{refs}

\end{document}

%% file: sections/01_introduction.tex
\section{Introduction}
\label{sec:intro}

Modern LLM agents now issue \emph{batches} of semantically related vector queries.
Query rewriting sends $N$ paraphrases of a question in parallel~\cite{queryrewrite,query2doc} (Figure~\ref{fig:teaser}).
Deep-research agents decompose a task into concurrent retrieval sub-tasks. Anthropic's Research system spins up 3--5 subagents in parallel and the subagents use 3+ tools in parallel~\cite{anthropic_multiagent}, and ParallelSearch~\cite{parallelsearch} trains the agent with RL to decompose a question and issue the sub-queries concurrently.
Post-training algorithms such as GRPO~\cite{grpo}, applied to search agents~\cite{searchr1}, sample multiple rollouts from the same prompt and produce semantically related searches.
HQI~\cite{hqi} reports that Apple's vector search workloads often need to be processed in a batch setting.

\begin{figure}[t]
  \centering
  \includegraphics[width=\columnwidth]{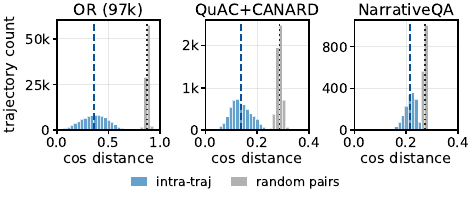}
  \caption{Pairwise cosine distance between queries of the same batch
  and between random query pairs.}
  \label{fig:intra-traj}
\end{figure}

In all of these patterns the batch members come from one intent and share semantic substance.
We measure this on three workloads, deep research (OpenResearcher~\cite{openresearcher}), conversational Q\&A (QuAC+CANARD~\cite{quac,canard}), and narrative reading comprehension (NarrativeQA~\cite{narrativeqa}).
Figure~\ref{fig:intra-traj} shows that mean intra-batch cosine distance on these workloads is $2.4{\times}$, $2.1{\times}$, and $1.3{\times}$ tighter than random query pairs respectively.

Existing disk-resident ANN systems process every query on its own.
Queries that are close search the same part of the index, so a batch reads the same data from SSD again and again.
With per-query \pipea~\cite{pipeann} at recall@10 of $0.95$, $67\%$, $31\%$, and $38\%$ of the I/Os of a batch on the three workloads fetch a page that a batch-mate already fetched (Sec.~\ref{sec:bg:repeat}).
A cluster-based index such as SPANN~\cite{spann} repeats $27\%$ to $68\%$ of its cluster I/Os on the same batches.

A cluster index knows its clusters before the first I/O and could deduplicate them across the batch, but it reads every full-precision vector of a cluster it probes.
A graph index reads one node at a time and selects it by a compressed distance held in memory.
Even if every query stops as early as the ground truth allows and the batch removes every repeated cluster, a cluster index reads $5.0$ to $7.2{\times}$ the data that per-query \pipea\ reads without any deduplication (Sec.~\ref{sec:bg:graph}).
We therefore build on a graph index.

On a graph index the direct fix is a page cache shared by the batch.
We build this baseline and call it GLRU.
It removes nearly all repeated I/Os, yet on one thread it is only $1.1$ to $1.2{\times}$ faster than \pipea\ (Sec.~\ref{sec:bg:cache}).
On one thread a search spends its time computing PQ and exact distances, because with up to $32$ I/Os in flight per query its I/Os complete while it computes.
A cache removes I/Os but leaves the computation unchanged.
A cached page serves only the queries whose walks ask for it, although the full-precision vector on it is a candidate result for every batch-mate.
When every query walks with a short candidate list, $29$ to $48\%$ of the true neighbors that its own walk misses lie on pages that a batch-mate read.
GLRU never ranks these pages for that query, so it needs the same candidate list length $L$ as \pipea\ and computes as much.

We present \sys, which runs the $B$ queries of a batch as cooperating walks on one thread (Sec.~\ref{sec:design}).
Every query keeps its own candidate list and selects its own next nodes, as a per-query search does.
The walks cooperate in three ways.
A page is read once per batch.
When a page arrives, \sys\ computes the exact distance from its node to every query of the batch, and at the end every query ranks all nodes the batch read.
A query therefore reaches the target recall with a smaller candidate list, fewer I/Os, and less computation.
A single \pipea\ query with $4$ I/Os in flight waits for I/Os $54$ to $73\%$ of its time.
One \sys\ thread keeps the I/Os of all $B$ walks in flight and processes whichever page completes first, so $4$ I/Os in flight per query keep it computing.
\sys\ does not share graph neighbors across queries.
PQ-scoring a fetched node's neighbors for the whole batch makes a batch $1.3$ to $2.4{\times}$ slower on one thread (Sec.~\ref{sec:eval:analysis}).

We evaluate \sys\ on the three workloads against \pipea\ and GLRU.
On one thread at recall@10 of $0.95$, \sys\ answers a batch $1.9{\times}$, $1.8{\times}$, and $1.6{\times}$ faster than \pipea\ and $1.8{\times}$, $1.8{\times}$, and $1.6{\times}$ faster than GLRU.
It issues $2.8$ to $6.9{\times}$ fewer I/Os than \pipea.
On a server, with every system at its knee thread count and each \sys\ batch on several threads, \sys\ serves $3.0$ to $6.9{\times}$ its throughput and $2.0$ to $3.0{\times}$ its queries per dollar while keeping the batch latency at $0.74$ to $1.08{\times}$ that of \pipea.

\paragraph{Contributions.} This paper makes four contributions.
\begin{itemize}
  \item We characterize agent batch retrieval on three workloads. We quantify the intra-batch similarity and the repeated I/Os it causes on both graph and cluster disk indexes, and we show why a graph index is the better base for batches (Sec.~\ref{sec:bg}).
  \item We show that a batch-scoped page cache removes the repeated I/Os but leaves most of the latency, and we identify what it leaves unused (Sec.~\ref{sec:bg:cache}).
  \item We design \sys, which runs a batch as cooperating walks on one thread that share pages and exact distances while every query keeps its own candidate list (Sec.~\ref{sec:design}).
  \item We evaluate \sys\ on three workloads, reaching $1.6$ to $1.9{\times}$ lower latency than \pipea\ at recall $0.95$, and we measure how much each form of cooperation and each rejected design alternative contributes (Sec.~\ref{sec:eval}).
\end{itemize}

The source code of \sys\ is available at \url{https://github.com/OhhhhPi/SWARM}.

%% file: sections/02_background_motivation.tex
\section{Background and Motivation}
\label{sec:bg}

\subsection{Disk-Resident ANN Indexes}
\label{sec:bg:ann}

\paragraph{Graph indexes.}
Disk-resident graph ANN indexes store the dataset as a Vamana
graph~\cite{diskann}. Each vector becomes a node with $R$ out-edges to
other vectors, and the graph is laid out on SSD so that one I/O returns
one node's full-precision vector together with the identifiers
of its $R$ neighbors. A compressed product-quantization (PQ) code
of every vector is held in DRAM and supports fast approximate distance
computation against any candidate node. A search query
maintains a length-$L$ candidate list sorted by PQ
distance and runs a beam search. At each hop the query pops up to
\texttt{beam\_width} unvisited nodes, fetches them from disk,
PQ-scores the fetched nodes' neighbors against its own lookup table,
and inserts the results back into the candidate list. The walk terminates when
no unvisited nodes remain, and the query ranks the nodes it fetched by
exact distance. \diska~\cite{diskann} introduced this
layout. \pipea~\cite{pipeann} pipelines the I/O and compute stages of the walk
so that SSD latency is overlapped with PQ scoring. We use \pipea\ as our
per-query baseline throughout.

\paragraph{Cluster indexes.}
Cluster-based indexes such as SPANN~\cite{spann} and SPFresh~\cite{spfresh}
partition the dataset into clusters. The centroids stay in memory, and
the full-precision vectors of each cluster are stored together on SSD.
A query ranks the centroids, reads the nearest clusters, and computes the
exact distance to every vector it read. The clusters to read are known
before the first I/O is issued. No compressed code filters the vectors
of a cluster, so probing a cluster reads all of them.

\subsection{Agent Batch Workloads}
\label{sec:bg:agents}

Six concrete agentic patterns drive the shift to batched related retrieval.
\emph{Query rewriting} generates $N$ paraphrases for a single question, hedging against embedding-space brittleness~\cite{queryrewrite,query2doc}.
\emph{Post-training algorithms} like GRPO~\cite{grpo}, when applied to memory or deep-research agents~\cite{searchr1}, sample multiple rollouts from the same prompt and naturally produce semantically related searches.
\emph{HyDE}~\cite{hyde} prompts an LLM to draft $N$ hypothetical answers and uses each answer's embedding as a separate query.
\emph{Sub-question decomposition} splits a complex question into atomic sub-questions, each issued as its own retrieval call~\cite{selfask,leasttomost}.
\emph{Plan-and-execute deep-research agents}, exemplified by Perplexity Deep Research, OpenAI Deep Research, and Anthropic Skill agents~\cite{openai_deepresearch,perplexity_deepresearch,anthropic_skills}, dispatch $5$ to $20$ retrieval sub-tasks in parallel per planning step.
\emph{Claim-grounded fact-checking} treats each generated sentence as a separate query for evidence grounding~\cite{factscore,safe}.
The shared property across all six patterns is that the batch members originate from one intent and vary in surface form rather than in semantic substance.
In our three evaluation workloads the mean batch size is $33.5$ (OpenResearcher), $7.2$ (QuAC+CANARD), and $29.7$ (NarrativeQA).

\paragraph{Batch members are close.}
\label{sec:bg:similarity}
If a batch genuinely comes from a single user intent, its queries
should be measurably closer in embedding space than random pairs drawn
from the same workload. We test this on the three datasets below.

\textbf{OpenResearcher}~\cite{openresearcher} is a deep-research-agent
benchmark. It provides $97{,}630$ multi-turn research trajectories,
each one an offline deep-research session run by a GPT-OSS-120B agent
and averaging $33.5$ retrievals issued as
\texttt{browser.search} tool calls. Its corpus consists of $14.9$M
FineWeb web-crawl passages.

\textbf{QuAC+CANARD} is a conversational QA workload over Wikipedia.
Each dialog is a sequence of follow-up questions that drill
into a single Wikipedia article. CANARD~\cite{canard} rewrites
each turn to stand on its own, so every one of the $5{,}644$ dialogs
becomes a batch of self-contained queries that retrieve independently yet
still probe the same article ($40{,}527$ queries in total).

\textbf{NarrativeQA}~\cite{narrativeqa} is a reading-comprehension
workload over $1{,}572$ books and movie scripts with $46{,}765$
human-written questions.

For OpenResearcher we use its official corpus embeddings produced
by Qwen3-Embedding-8B ($4096$d) and encode queries with the same
model. For QuAC+CANARD we use the $21$M-passage DPR Wikipedia corpus~\cite{dpr}
as the underlying collection. For NarrativeQA we chunk the original
corpus at $100$-word granularity. On QuAC+CANARD and NarrativeQA we
encode queries and passages with e5-base-v2 ($768$d).

In all three, every query is self-contained and independent of the other queries of its batch.
Figure~\ref{fig:intra-traj} shows that queries of the same batch are
systematically closer than random pairs on all three workloads.
Cross-panel absolute distances are not directly comparable, because OpenResearcher uses a different embedding model and dimensionality from the other two.
QuAC+CANARD and NarrativeQA share one embedding model, but their corpora differ, so even their mean random-pair distances differ slightly ($0.29$ vs $0.27$).
We therefore read each panel through the intra-versus-random \emph{gap}.
Intra-batch pairs lie $58.5\%$ closer than random pairs on OpenResearcher ($0.36$ vs $0.87$), $52.3\%$ closer on QuAC+CANARD ($0.14$ vs $0.29$), and $21.0\%$ closer on NarrativeQA ($0.22$ vs $0.27$).
NarrativeQA's smaller gap is structural, because each batch
groups independent reading-comprehension questions about different
characters and scenes in one narrative.

\subsection{Per-Query Search Repeats I/Os}
\label{sec:bg:repeat}
\label{sec:bg:joint-pays}

\begin{figure}[t]
  \centering
  \includegraphics[width=\columnwidth]{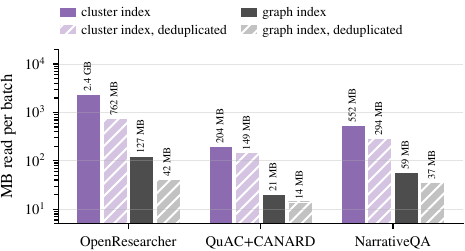}
  \caption{Data read from SSD per batch, with the graph index at recall@10 of $0.95$ and the cluster index at a lower bound.}
  \label{fig:batch-bytes}
\end{figure}

Queries that are close search the same part of an index.
On a graph index their walks start from the same entry point and expand many of the same nodes.
On a cluster index they rank many of the same centroids first.
A system that processes each query on its own issues these I/Os once per query.

We count the repeated I/Os on both index families on the first $1{,}000$ batches of each workload.
For the graph index we run \pipea\ per query at recall@10 of $0.95$ and give each batch a page cache that never evicts.
The cache misses of a batch are then its distinct pages and the hits are its repeated I/Os.
For the cluster index we measure offline and build nothing on disk.
We partition each corpus with $k$-means into clusters of $64$ vectors on average, replicate boundary vectors into nearby clusters as SPANN does, and rank the centroids for every query.
A cluster index ranks the vectors it reads by exact distance, so the recall of a query is the fraction of its true top-$10$ that lies in the clusters it probes.
We let every query stop as early as the ground truth allows.
A query probes clusters in centroid order until $9$ of its true top-$10$ are found, but never more than $344$, $715$, and $369$ clusters on the three workloads, the single depth at which the mean recall over all queries reaches $0.95$.
The mean recall is then $0.89$ to $0.90$, below the $0.95$ of \pipea.
For each batch we then compare the clusters its queries request against the distinct clusters among them.

Figure~\ref{fig:batch-bytes} shows the data a batch reads with and without the repeats.
On the graph index $67\%$, $31\%$, and $38\%$ of the I/Os of a batch on OpenResearcher, QuAC+CANARD, and NarrativeQA fetch a page that a batch-mate already fetched.
On OpenResearcher the share grows with the batch, from $40\%$ for $B{\in}[2,6]$ to $69\%$ for $B{\in}[60,102]$.
On the cluster index $68\%$, $27\%$, and $47\%$ of the cluster I/Os are repeats.
The repeats come from the workload, and both index families pay for them.

\subsection{Graph or Cluster Index}
\label{sec:bg:graph}

Even after a batch removes every repeated I/O, a cluster index has worse lower bounds on batch latency and throughput than per-query \pipea.
If it keeps only its centroids in memory, as SPANN does, it reads every full-precision vector of a probed cluster from SSD.
Even if each query stops as soon as the ground truth allows, after $72$ to $146$ clusters on average, a deduplicated batch reads $5.0$ to $7.2{\times}$ the data of \pipea\ without deduplication (Figure~\ref{fig:batch-bytes}), and reading it at the $11.0$\,GB/s of one thread takes $1.9$ to $7.0{\times}$ the batch latency of \pipea.

If it keeps PQ codes in memory instead and reranks only the best candidates from SSD, its recall still requires the probed clusters to contain the true neighbors, so on OpenResearcher every query probes about $1{,}500$ clusters and scans about $150$K codes to reach a mean recall of $0.95$.
Before any I/O, a batch then takes at least $6.3{\times}$ the single-thread latency of \pipea\ with $8$-bit codes and $1.2$ to $1.7{\times}$ with $4$-bit codes.
Queries that probe the same cluster read the same codes but compute their own distances, so ranking the candidates reranked by the whole batch still leaves $1.1$ to $1.4{\times}$.

A graph index avoids both costs because its edges lead toward the query, so it scores only the neighbors of the nodes it expands and reads only the nodes among its $L$ best.
On OpenResearcher \pipea\ reaches recall $0.95$ with about $190$ I/Os and $12{,}500$ PQ distances per query.
We therefore build \sys\ on a graph index.

\subsection{A Page Cache Is Not Enough}
\label{sec:bg:cache}

On a graph index the direct way to remove the repeated I/Os is a page cache shared by the queries of a batch.
We build this baseline on \pipea\ and call it GLRU (Sec.~\ref{sec:eval:baselines}).
On one thread it reads every distinct page of a batch exactly once.
Yet on one thread GLRU answers a batch only $1.1$ to $1.2{\times}$ faster than \pipea\ (Sec.~\ref{sec:eval:analysis}).

On one thread the search time is computation.
For every fetched node a query computes the exact distance of the node and the PQ distances of its neighbors, and the I/Os of later nodes complete in the meantime.
\pipea\ overlaps I/O and computation by keeping several I/Os in flight per query.
With $32$ I/Os in flight it only waits for I/Os $16\%$, $13\%$, and $21\%$ of its time on the three workloads.
A cache that removed all of this waiting would make a batch at most $1.19$, $1.15$, and $1.26{\times}$ faster.

A cache therefore removes only I/Os whose latency is already hidden, and it leaves the computation unchanged.
The computation is set by the path of the walk, because the path decides which nodes a query fetches and how many distances it computes.
The cache is passive and serves only the I/Os the walk has already decided to issue, so it does not change the path.
As a result GLRU reaches the same recall as \pipea\ at the same $L$ and computes as many distances.
We measure this gain in Sec.~\ref{sec:eval:analysis}.

The cache leaves the full-precision vectors on its pages unused.
We let every query of a batch walk on its own with a short candidate list, $L$ of $70$, $304$, and $248$, while the batch reads each page once.
The walk of a query then misses $9.3\%$, $8.5\%$, and $7.7\%$ of its true top-$10$, and $48\%$, $29\%$, and $38\%$ of the missed neighbors lie on pages that a batch-mate read.
These pages are already in memory, but a cache returns each page only to the walks that select it, so the query never ranks these neighbors.

%% file: sections/03_design.tex
\section{\sys\ Design}
\label{sec:design}

\subsection{Overview}
\label{sec:design:overview}

\sys\ answers a batch of $B$ queries on one thread with the same disk-resident Vamana index, PQ codes, and parameters as \pipea.
It runs one walk per query.
Each walk keeps its own candidate list of length $L$ sorted by PQ distance, selects its own next nodes from that list, keeps up to four I/Os pending, and stops when its list holds no unexpanded node and none of its I/Os is pending.
A query's walk therefore selects nodes by the same rule as a per-query search, from a list which is only filled by its own expansions.
A batch of one query has no batch-mate, and its walk would keep only $4$ I/Os in flight, so \sys\ runs it with \pipea.

\begin{figure}[t]
  \centering
  \includegraphics[width=\columnwidth]{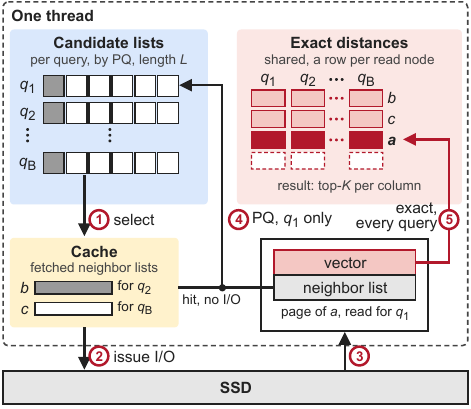}
  \caption{\sys\ on one thread.}
  \label{fig:overview}
\end{figure}

The walks share three things (Figure~\ref{fig:overview}).
They share I/Os, so a page that several walks select is read once (Sec.~\ref{sec:design:pages}).
They share exact distances, so every query ranks every node that any walk of the batch read (Sec.~\ref{sec:design:exact}).
They share one thread, which keeps the I/Os of all walks in flight and processes whichever page completes first (Sec.~\ref{sec:design:thread}).
The walks do not share the neighbors they discover, and Sec.~\ref{sec:design:noshare} explains why.
Figure~\ref{fig:overview} follows the thread through one round.
Each query selects the best node from its candidate list~\stepnum{1}.
If the batch has already read the selected node, \sys\ expands it at once from the neighbor list it keeps in its cache, as $q_1$ does for node $b$, which was read for $q_2$.
Otherwise the thread issues an I/O for the node's page~\stepnum{2}.
When an I/O completes~\stepnum{3}, \sys\ PQ-scores the neighbors of its node for the queries that selected it and inserts them into their candidate lists~\stepnum{4}.
The first time a node is read, \sys\ also adds a row for it to the batch's exact-distance table, with its exact distance to every query~\stepnum{5}.
Each query returns the $K$ nodes with the smallest exact distance in its column of the table.
Algorithm~\ref{alg:cowalk} states the whole procedure.

\begin{algorithm}[t]
\caption{\sys\ for a batch $q_1, \ldots, q_B$ on one thread.}
\label{alg:cowalk}
\begin{algorithmic}[1]
\Require list length $L$, I/Os per query $w{=}4$
\Ensure top-$K$ of every query
\State \textbf{for each} $q_i$: $C_i \gets \{\textit{entry point}\}$
\State $E \gets$ an empty table with one column per query
\State \Call{Issue}{$q_i$} for every $q_i$; submit the queued I/Os
\While{some I/O is pending}
  \State wait until at least one I/O completes
  \ForAll{completed I/Os of a page $p$}
    \State \Call{Expand}{$p$, the queries that selected nodes on $p$}
  \EndFor
  \State \Call{Issue}{$q_i$} for every walking $q_i$; submit
  \ForAll{$n$ first read in this round}
    \State add to $E$ the row of exact distances of $n$
  \EndFor
\EndWhile
\State \Return the top-$K$ of each column of $E$
\Statex
\Procedure{Issue}{$q_i$}
  \While{$q_i$ has fewer than $w$ pending I/Os}
    \State $n \gets$ the best unexpanded node of $C_i$
    \State \textbf{if} none, stop $q_i$ when nothing is pending; \Return
    \If{the page of $n$ is being read}
      \State add $q_i$ to the queries waiting for that page
    \ElsIf{the page of $n$ has been read in this batch}
      \State \Call{Expand}{that page, $\{q_i\}$}
    \Else
      \State queue an I/O for the page for $q_i$
    \EndIf
  \EndWhile
\EndProcedure
\Statex
\Procedure{Expand}{$p$, $P$}
  \ForAll{$q_i \in P$}
    \State PQ-score the neighbors of the node on $p$ that $q_i$ has not scored
    \State insert them into $C_i$
  \EndFor
\EndProcedure
\end{algorithmic}
\end{algorithm}

\subsection{Sharing I/Os}
\label{sec:design:pages}

When a walk selects a node, \sys\ looks up the page that holds it.
If another walk has already issued an I/O for that page, the query joins that I/O and issues nothing.
If the page has already been read in this batch, \sys\ expands the node for the query at once.
Only otherwise does it issue a new I/O.
When an I/O completes, \sys\ copies the neighbor list of its node into a record that stays in memory until the batch ends (Sec.~\ref{sec:impl:records}).
A page is therefore read at most once per batch.

Expanding a node works as in a per-query search, but only for the queries that selected it.
For each such query, \sys\ computes the PQ distance to every neighbor of the node and inserts the neighbor into that query's candidate list if it ranks among the top $L$.
Other queries do not see these neighbors.
Each query skips the neighbors it has already PQ-scored, because a second score gives the same distance and cannot change its list.

\subsection{Sharing Exact Distances}
\label{sec:design:exact}

A page holds the full-precision vector of a node.
The first time any walk reads the node, \sys\ computes the exact distance from that vector to \textbf{every query of the batch} and stores the distances as one row of the batch's exact-distance table, which has one column per query.
This includes the queries whose walks have already stopped, because a node that a batch-mate reads later can still be one of their nearest neighbors.
One kernel loads the vector once and scores it against four queries at a time (Sec.~\ref{sec:impl:simd}).
No query needs the vector after this step.
Each query returns the $K$ nodes with the smallest exact distance in its column.

Every query thus ranks every node that the batch read.
A per-query search ranks only the nodes its own walk read, so a true neighbor enters its result only if its own walk reaches that node, and only a longer candidate list can make the walk reach it.
In \sys\ a true neighbor of a query that a batch-mate read is already in the query's column of the table.
The query reaches the same recall with a shorter candidate list, which expands fewer nodes and so issues fewer I/Os and fewer PQ comparisons (Sec.~\ref{sec:eval:analysis}).

Exact distances only rank.
They never enter a candidate list, so a batch-mate's I/Os change which nodes a query ranks at the end but not which nodes its walk selects.
A query far from the rest of its batch is therefore never pulled toward the region its batch-mates search.
The price is one exact distance per read node and query of the batch, which grows with $B$.

\subsection{One Thread for All Walks}
\label{sec:design:thread}

\begin{figure}[t]
  \centering
  \includegraphics[width=\columnwidth]{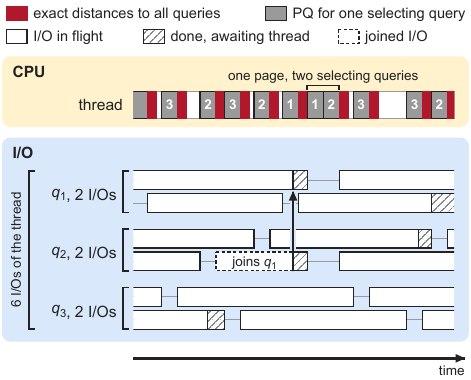}
  \caption{One \sys\ thread serving three queries, with two I/Os per query shown for clarity. Durations are illustrative.}
  \label{fig:pipeline}
\end{figure}

All walks issue their I/Os to one \texttt{io\_uring} instance on the thread.
The thread waits until at least one I/O completes and expands the node of every completed page for the queries that selected it.
It then lets every walking query with fewer than four pending I/Os select more nodes, and it submits all I/Os queued in this round with one system call.
Only then does it compute the exact distances of the nodes read for the first time in this round, so the new I/Os are in flight while it does.
The exact distances never enter a candidate list, so computing them after the submission leaves every selection unchanged.

Figure~\ref{fig:pipeline} shows this loop for three queries with two I/Os each.
On the CPU lane a grey block is the PQ work for one query that selected a page, and a red block is the exact distances from the page's node to the whole batch.
A page is expanded as soon as it completes, whichever walk it belongs to, so the I/Os of one query complete while the thread works on the pages of the others.
A completed I/O waits (hatched) only while the thread is busy, and the thread blocks only when none of the I/Os of the $B$ walks has completed.
When $q_2$ selects a node whose page $q_1$ is already reading, $q_2$ joins that I/O, and the page gets a grey block for each of them and one red block.
With $B$ walks the thread has up to $4B$ I/Os in flight ($2B$ in Figure~\ref{fig:pipeline} for clarity), so four per query keep it computing where a single \pipea\ query needs up to $32$ (Sec.~\ref{sec:bg:cache}).
More I/Os per query would not help, because a query selects all of its pending I/Os from the same candidate list, before their neighbors arrive, so each extra I/O is more likely to fetch a node the query would no longer select (Sec.~\ref{sec:eval:analysis}).

The alternative is a walk in synchronous hops.
Every query issues its I/Os for the hop, and the thread waits for all of them before it processes any page.
One slow I/O then holds up every walk of the batch, and the thread idles while the last I/Os of a round are in flight.
Sec.~\ref{sec:eval:analysis} runs this schedule with the same sharing.

\subsection{What \sys\ Does Not Share}
\label{sec:design:noshare}

A read page holds both the vector of a node and its neighbor list.
Both arrive with the same I/O, so a natural design would also share the neighbor list and PQ-score its neighbors for every query of the batch.
\sys\ shares only the vector.
The query that selected a node ranked it among its $L$ best by distance to its own vector, so the neighbors of the node are the most promising candidates that query has.
A batch-mate receives these neighbors through a node ranked by distance to another query vector, and even its nearest batch-mate lies at some distance from it, so they are less promising for it than the neighbors of the nodes it selected itself.
In an earlier hop-synchronous version of the walk that scored every neighbor for the whole batch, $16\%$, $14\%$, and $10\%$ of the neighbors of a query's own nodes entered its candidate list, but only $5.6\%$, $7.3\%$, and $6.0\%$ of the neighbors of the nodes its nearest batch-mate selected, and $3.5\%$, $5.6\%$, and $4.1\%$ excluding neighbors that its own nodes also list.
Scoring them for the whole batch would add $B$ PQ distances per neighbor, and PQ is already the largest stage of the walk.
The value of an exact distance does not depend on which query selected the node.
It costs one distance per query and read node, and it ranks the node exactly for every query of the batch (Sec.~\ref{sec:design:exact}).
Sec.~\ref{sec:eval:analysis} measures a variant that scores the neighbors for the whole batch.

%% file: sections/04_implementation.tex
\section{Implementation}
\label{sec:impl}

We implement \sys\ in the code base of \pipea\ and reuse its index format, PQ distance code, and \texttt{io\_uring} reader.

\paragraph{Cross-query SIMD kernel.}\label{sec:impl:simd}
The exact distances of a node go to every query of the batch.
\sys\ computes them with an AVX-512 kernel that loads each block of the node vector once and multiplies it into four accumulators, one per query.
The vector is thus read once per four queries, and the four independent fused multiply-adds overlap in the pipeline.

\paragraph{Scored neighbors.}\label{sec:impl:layout}
Each query records the neighbors it has PQ-scored in an array of node ids indexed by a hash of the id, with at least $2LR$ entries for out-degree $R$.
A collision overwrites an entry, which can only make the query score a neighbor again and never drops a candidate.

\paragraph{Node records.}\label{sec:impl:records}
When an I/O completes, \sys\ copies the neighbor list of the node into a record that lives until the batch ends, and the I/O buffer returns to the pool.
A hash table indexed by page id points to the records, so a later pick of the node by any query of the batch expands it without an I/O.

\paragraph{Several threads per batch.}\label{sec:impl:team}
With $k$ threads per batch, the queries are dealt round-robin to the threads, which share the page table through atomic updates without locks.
A thread enters a page in the table when it issues the I/O, so another thread that picks the page waits for that I/O and does not read it again.
During the walk a thread computes exact distances only for its own walking queries.
After all walks end it fills in the rest, including the nodes the other threads read, so the records also keep the node's vector.

%% file: sections/05_evaluation.tex
\section{Evaluation}
\label{sec:eval}

\begin{table}[t]
  \caption{Evaluation workloads. Tighter: intra- vs.\ random-pair
  cosine gap (Sec.~\ref{sec:bg:similarity}); $\bar{B}$: mean batch size.}
  \label{tab:datasets}
  \centering
  \resizebox{\columnwidth}{!}{%
  \begin{tabular}{lrlrrr}
    \toprule
    Workload & $N{\times}d$ & Embedding & Batches & $\bar{B}$ & Tighter \\
    \midrule
    OpenResearcher & 14.9M$\times$4096 & Qwen3-8B & 97{,}630 & 33.5 & $58.5\%$ \\
    QuAC+CANARD    & 21M$\times$768    & e5-base  & 5{,}644  & 7.2 & $52.3\%$ \\
    NarrativeQA    & 0.77M$\times$768  & e5-base  & 1{,}572  & 29.7 & $21.0\%$ \\
    \bottomrule
  \end{tabular}}
\end{table}

We evaluate \sys\ against the strongest per-query baselines on three
application-realistic workloads, organized around three questions.
\begin{itemize}
  \item How much faster is \sys\ end to end, on a personal device and on a server (Sec.~\ref{sec:eval:main})?
  \item Where does the speedup come from, and how much does each design choice contribute (Sec.~\ref{sec:eval:analysis})?
  \item How would throughput grow with more SSDs (Sec.~\ref{sec:eval:scaling})?
\end{itemize}

\subsection{Experimental Setup}
\label{sec:eval:setup}

\begin{table}[t]
  \caption{Index and search parameters.}
  \label{tab:params}
  \centering
  \resizebox{\columnwidth}{!}{%
  \begin{tabular}{ll}
    \toprule
    Parameter & Value \\
    \midrule
    Graph max out-degree $R$   & $128$ (OR), $64$ (others) \\
    Build candidate list $L_b$ & $256$ (OR), $128$ (others) \\
    PQ code                    & $192$\,B/vector (OR), $96$\,B (others) \\
    Beam width (I/Os in flight per query)  & \sys\ $4$; \pipea, GLRU $32$ \\
    \bottomrule
  \end{tabular}}
\end{table}

\begin{figure}[t]
  \centering
  \includegraphics[width=\columnwidth]{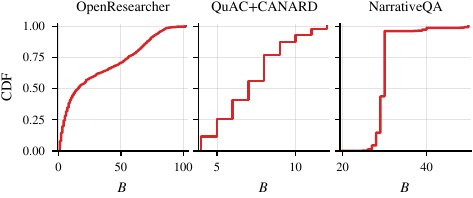}
  \caption{Batch-size ($B$) distribution per workload.}
  \label{fig:bsize-cdf}
\end{figure}

\paragraph{Hardware.}
We run every experiment on a single server:
\begin{itemize}
  \item \textbf{CPU}: dual-socket AMD EPYC 9754 (Zen 4c, $256$ cores / $512$ threads, two NUMA nodes, AVX-512). We pin every run to one NUMA node with \texttt{numactl};
  \item \textbf{Memory}: $1.5$\,TB DDR5 ($24{\times}64$\,GB, $4800$\,MT/s);
  \item \textbf{SSD}: Huawei eKitStor Xtreme 116P ($30.72$\,TB, PCIe 5.0 ${\times}4$);
  \item \textbf{OS}: Ubuntu $24.04$ with Linux kernel $6.8$.
\end{itemize}

\paragraph{Datasets.}
Table~\ref{tab:datasets} lists the three workloads, which span three application domains and differ widely in scale, embedding dimension, and intra-batch tightness. 
Figure~\ref{fig:bsize-cdf} shows their batch-size distributions.
Only OpenResearcher has batches of one query, $8\%$ of the batches we run and $0.3\%$ of their queries.
\sys\ runs them with \pipea\ (Sec.~\ref{sec:design:overview}), so we report recall and single-thread latency over the batches with $B{\geq}2$.

\paragraph{Parameters.}
Table~\ref{tab:params} lists the index and search parameters. 
All systems share one disk-resident Vamana index per workload and build no auxiliary structures. 

\paragraph{Baselines.}\label{sec:eval:baselines}
We compare against two per-query baselines on the same index and I/O engine. 
\begin{itemize}
  \item \emph{\pipea}~\cite{pipeann} is the strongest per-query system and our main point of comparison. 
  \item \emph{GLRU} runs \pipea\ and gives the queries of a batch, which run on different threads, one shared page cache of $32{,}768$ pages.
  The cache is emptied when the batch ends, so every hit comes from a page that another query of the same batch has read.
  At recall $0.95$ no batch of the three workloads evicts a page.
\end{itemize}
We test no other cache policy because GLRU is already at the ceiling of caching. 
On one thread GLRU reads every distinct page of a batch exactly once, a floor that no cache can beat.
On $128$ threads it reads $12\%$, $2.5\%$, and $2.3\%$ more, because two queries of a batch that run at the same time can both miss a page before its first I/O completes.

We omit Starling~\cite{starling} because a single $4096$-d OpenResearcher vector is $16$\,KB and a $768$-d vector on our other two workloads is $3$\,KB, so no $4$\,KB page holds more than one node and the optimization does not apply.
We also omit cluster-based disk indexes such as SPANN~\cite{spann}, which read more data or compute more distances per batch than a graph index at the same recall (Sec.~\ref{sec:bg:graph}).

\paragraph{Scenarios.}
We evaluate two deployments at recall@10 of $0.90$ and $0.95$.
\paragraph{Personal device.} 
A personal knowledge base shares a laptop or workstation with the model, its runtime, and whatever the model offloads to memory. 
Cores and memory are scarce on these devices, and retrieval must keep latency low and recall high  while using few resources to keep user experience.
Throughput matters less in this scenario. 
We give every system one thread and measure the time from a batch's arrival until all $B$ queries complete as latency. 
\paragraph{Server.} 
An online service cares more about throughput and cost. 
Retrieval is a small part of a request whose time goes mostly to post-processing and LLM decoding, so we care more about the QPS served per unit of cost.
We run each system at its knee, the smallest thread count that reaches $95\%$ of its peak QPS.
A \pipea\ thread serves one query at a time, and \sys\ runs each batch on $k$ threads (Sec.~\ref{sec:eval:main}).

\subsection{End-to-End Performance}
\label{sec:eval:main}

\begin{figure}[t]
  \centering
  \includegraphics[width=\columnwidth]{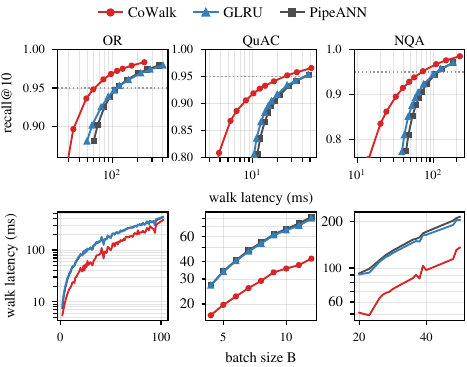}
  \caption{Single-thread walk latency. Top: recall@10 vs.\ latency. Bottom: latency vs.\ $B$ at recall $0.95$.}
  \label{fig:pareto-latency}
\end{figure}

\begin{figure}[t]
  \centering
  \includegraphics[width=\columnwidth]{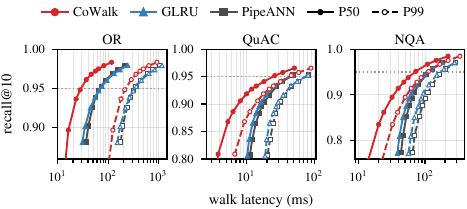}
  \caption{Single-thread walk latency percentiles over batches vs.\ recall@10.}
  \label{fig:tail-latency}
\end{figure}

\begin{figure}[t]
  \centering
  \includegraphics[width=\columnwidth]{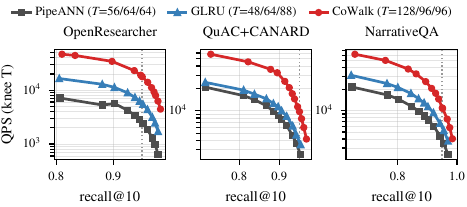}
  \caption{Throughput at each system's knee thread count vs.\ recall@10.}
  \label{fig:pareto-qps}
\end{figure}

\paragraph{Personal device.}
Figure~\ref{fig:pareto-latency} shows single-thread latency.
In the top row each curve sweeps $L$, and each point is the mean recall and mean latency over all batches with $B{\geq}2$ at one $L$.
A curve further to the left reaches the same recall in less time.
The bottom row plots the mean latency of the batches of each size $B$, with every system at the smallest $L$ of its sweep that reaches recall $0.95$.

At recall $0.95$ \sys\ answers a batch in $62$, $25$, and $71$\,ms on OpenResearcher, QuAC+CANARD, and NarrativeQA.
\pipea\ takes $115$, $45$, and $116$\,ms, which is $1.9{\times}$, $1.8{\times}$, and $1.6{\times}$ longer, and GLRU takes $1.8{\times}$, $1.8{\times}$, and $1.6{\times}$ longer.
At recall $0.90$ the gap to \pipea\ is $1.9{\times}$, $2.3{\times}$, and $2.0{\times}$.
Figure~\ref{fig:tail-latency} replaces the mean over batches with the P50 and the P99.
\sys's P50 at recall $0.95$ is $29$, $25$, and $71$\,ms, $2.4{\times}$, $1.8{\times}$, and $1.6{\times}$ below \pipea's and $2.4{\times}$, $1.8{\times}$, and $1.6{\times}$ below GLRU's.
Its P99 is $224$, $43$, and $111$\,ms, $1.5{\times}$, $1.8{\times}$, and $1.7{\times}$ below \pipea's and $1.5{\times}$, $1.7{\times}$, and $1.6{\times}$ below GLRU's.
\sys\ also reaches higher recall.
In the bottom row the speedup over \pipea\ grows from $1.6{\times}$ at $B{=}4$ to $2.0{\times}$ at $B{=}12$ on QuAC+CANARD and stays near $1.8{\times}$ on NarrativeQA.
On OpenResearcher it is $1.4{\times}$ at $B{=}2$, $2.0{\times}$ on average for $B$ from $10$ to $30$, and $1.4{\times}$ for $B{\geq}60$.

On one thread I/O is not the bottleneck.
With $32$ I/Os in flight a \pipea\ walk waits on I/Os for only $13$ to $21\%$ of its time, and the other I/Os complete while it computes.
All three systems therefore spend most of their time computing PQ and exact distances, and this computation grows with $L$.
GLRU walks with the same $L$ as \pipea\ and only skips the repeated I/Os, so its curves almost overlap those of \pipea.
\sys\ is faster because every query also ranks the nodes its batch-mates fetched.
It reaches recall $0.95$ with $L$ of $70$, $304$, and $248$, against $184$, $688$, and $464$ for \pipea.
Its $B$ walks keep the thread computing with $4$ I/Os in flight per query, where \pipea\ and GLRU need up to $32$.
Sec.~\ref{sec:eval:analysis} measures both.

On OpenResearcher all three curves in the bottom row are jagged.
Each point averages only the batches of one exact size, and half of the sizes have at most six batches.
Batches of the same size take different times per query, and so few batches do not average these differences out.

\paragraph{Server.}
Figure~\ref{fig:pareto-qps} shows throughput with every system at its knee, the smallest thread count that reaches $95\%$ of its peak QPS.
\pipea\ and GLRU hand the queries of a batch to different threads, and their batch latency is the time from the start of the batch's first query to the end of its last.
\sys\ runs a batch on $k$ threads and $T/k$ batches at a time (Sec.~\ref{sec:impl:team}), and its batch latency is the time from the start of the batch until the last of its $k$ threads finishes the exact distances.
We use about four queries per thread of an average batch.
The server runs use the first $4{,}000$ batches of OpenResearcher and all batches of the other two, whose mean sizes of $29.5$, $7.2$, and $29.7$ give $k{=}8$, $2$, and $8$.
In these runs the batches of one query, $0.3\%$ of the OpenResearcher queries, still run the walk of \sys.
Figure~\ref{fig:pareto-qps} runs \sys\ with this $k$ at its knee of $128$, $96$, and $96$ threads.

At recall $0.95$ on the thread ladder of Figure~\ref{fig:tladder}, \sys\ serves $17.4$K, $10.0$K, and $11.1$K QPS, which is $6.9{\times}$, $4.0{\times}$, and $3.0{\times}$ the throughput of \pipea\ and $3.0{\times}$, $2.8{\times}$, and $1.9{\times}$ that of GLRU.
At recall $0.90$ it serves $6.0{\times}$, $3.7{\times}$, and $3.2{\times}$ the throughput of \pipea.
Its batch latency at the knee is $28$, $34$, and $32$\,ms, $0.74$ to $1.08{\times}$ the $37$, $32$, and $29$\,ms of \pipea.
GLRU reaches a lower batch latency of $15$, $22$, and $24$\,ms at its knee.

Figure~\ref{fig:tladder} places every thread count of every system at its QPS and batch latency, so it compares the systems at equal latency.
We take the \sys\ configuration, over all $k$ and thread counts, with the highest QPS whose batch latency does not exceed that of a baseline at its knee.
Within the batch latency of \pipea\ it serves $6.9{\times}$, $3.6{\times}$, and $2.6{\times}$ the QPS of \pipea, with $k$ of $8$, $4$, and $8$.
Within the batch latency of GLRU it serves $2.6{\times}$, $2.5{\times}$, and $1.7{\times}$ the QPS of GLRU, with $k$ of $16$, $4$, and $8$.

With $k{=}1$ \sys\ reaches its highest throughput, $17.9$K, $10.1$K, and $12.9$K QPS at recall $0.95$, so the $k$ above keeps $97\%$, $99\%$, and $86\%$ of it.
When threads are added step by step (Figure~\ref{fig:tladder}), \sys\ with $k{=}1$ on $8$ threads (OpenResearcher) or $16$ threads (the other two) already exceeds the peak that \pipea\ reaches on $128$.
Its batch latency is $62$\,ms with one batch at a time and $66$\,ms with $16$ on OpenResearcher, and grows to $97$, $45$, and $126$\,ms at the knee, $3.5{\times}$, $1.3{\times}$, and $4.0{\times}$ that of the $k$ above.
Sec.~\ref{sec:eval:analysis} measures other values of $k$.

With many threads the SSD limits \sys, so its throughput is the I/Os the SSD completes per second divided by the I/Os a query needs.
\sys\ issues $2.8$ to $6.9{\times}$ fewer I/Os per query than \pipea, because batch-mates read each page once and exact-distance sharing lowers $L$.
Four I/Os in flight per query instead of $32$ also contribute, because at equal $L$ GLRU, which also reads each page once per batch, issues $7$ to $10\%$ more I/Os than \sys.
Sec.~\ref{sec:eval:analysis} measures the two effects.

\begin{table}[t]
  \caption{Cost at the knee (recall $0.95$, Google Cloud on-demand prices). Mq/\$: million queries per dollar. Ratio: Mq/\$ over \pipea.}
  \label{tab:cost}
  \centering
  \setlength{\tabcolsep}{3.5pt}
  \begin{tabular}{lrrrrrr}
    \toprule
    System & Threads & GiB & \$/h & KQPS & Mq/\$ & Ratio \\
    \midrule
    \multicolumn{7}{l}{\emph{OpenResearcher}} \\
    \pipea          & 56  & 3.5  & 1.71 & 2.5  & 5.4  & $1.0{\times}$ \\
    GLRU            & 48  & 5.2  & 1.48 & 5.7  & 14.0 & $2.6{\times}$ \\
    \sys\ $k{=}1$   & 56  & 5.0  & 1.72 & 17.9 & 37.6 & $7.0{\times}$ \\
    \sys\ $k{=}8$   & 128 & 4.6  & 3.84 & 17.4 & 16.3 & $3.0{\times}$ \\
    \midrule
    \multicolumn{7}{l}{\emph{QuAC+CANARD}} \\
    \pipea          & 64  & 3.0  & 1.95 & 2.5  & 4.7  & $1.0{\times}$ \\
    GLRU            & 64  & 3.5  & 1.95 & 3.6  & 6.6  & $1.4{\times}$ \\
    \sys\ $k{=}1$   & 64  & 3.0  & 1.95 & 10.1 & 18.8 & $4.0{\times}$ \\
    \sys\ $k{=}2$   & 96  & 3.3  & 2.89 & 10.0 & 12.4 & $2.7{\times}$ \\
    \midrule
    \multicolumn{7}{l}{\emph{NarrativeQA}} \\
    \pipea          & 64  & 0.8  & 1.94 & 3.7  & 6.9  & $1.0{\times}$ \\
    GLRU            & 88  & 1.4  & 2.65 & 5.7  & 7.8  & $1.1{\times}$ \\
    \sys\ $k{=}1$   & 56  & 1.2  & 1.70 & 12.9 & 27.4 & $4.0{\times}$ \\
    \sys\ $k{=}8$   & 96  & 0.8  & 2.88 & 11.1 & 13.8 & $2.0{\times}$ \\
    \bottomrule
  \end{tabular}
\end{table}

\paragraph{Cost.}
Table~\ref{tab:cost} prices every system at its knee in Figure~\ref{fig:tladder}.
We use the on-demand rates of Google Cloud C2D instances in us-central1, $\$0.0296$ per vCPU-hour and $\$0.0040$ per GiB-hour of memory, and $\$0.08$ per GiB-month of Local SSD.
Each thread takes one vCPU.
Memory is the peak resident memory of the index and the search at the knee, which Sec.~\ref{sec:eval:analysis} measures.
Every index fits on one $375$\,GiB Local SSD.
vCPUs account for $96\%$ to $98\%$ of the hourly cost, so the queries served per dollar follow the QPS per thread at the knee.
With the $k$ of Figure~\ref{fig:pareto-qps} \sys\ serves $3.0{\times}$, $2.7{\times}$, and $2.0{\times}$ the queries per dollar of \pipea, against $2.6{\times}$, $1.4{\times}$, and $1.1{\times}$ for GLRU.
With $k{=}1$ its knee needs $56$ to $64$ threads, and the ratio rises to $7.0{\times}$, $4.0{\times}$, and $4.0{\times}$.

\subsection{Analysis}
\label{sec:eval:analysis}

\begin{figure}[t]
  \centering
  \includegraphics[width=\columnwidth]{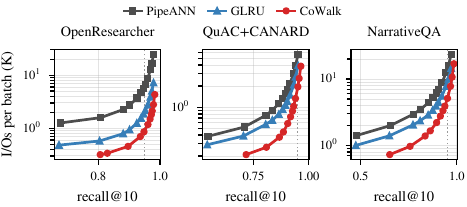}
  \caption{I/Os per batch vs.\ recall@10.}
  \label{fig:mean-io}
\end{figure}

\begin{figure}[t]
  \centering
  \includegraphics[width=\columnwidth]{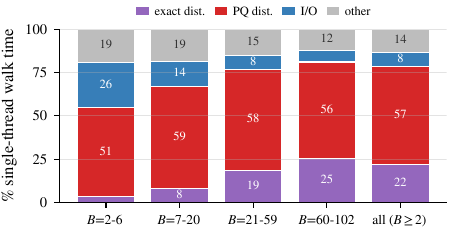}
  \caption{Single-thread walk time by stage and batch-size bucket (OpenResearcher).}
  \label{fig:stage-profile}
\end{figure}

\begin{figure}[t]
  \centering
  \includegraphics[width=\columnwidth]{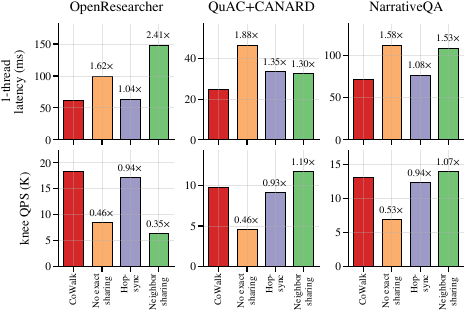}
  \caption{Single-thread batch latency and knee throughput of \sys\ and three variants at recall $0.95$.}
  \label{fig:variants}
\end{figure}

\begin{figure}[t]
  \centering
  \includegraphics[width=\columnwidth]{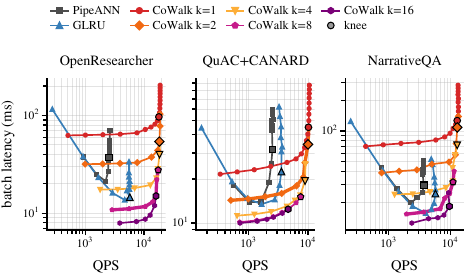}
  \caption{Batch latency vs.\ throughput at recall $0.95$ per total thread count, with $k$ threads per \sys\ batch (thick line: the $k$ chosen in the text).}
  \label{fig:tladder}
\end{figure}

This section explains the results of Sec.~\ref{sec:eval:main}.
It first counts I/Os.
It then changes one design choice of \sys\ at a time and measures latency and throughput again.
Last, it varies the I/Os in flight per query and the threads per batch.

\paragraph{I/O count.}
Figure~\ref{fig:mean-io} plots the I/Os of a batch against recall.
Each curve sweeps $L$, and a lower curve issues fewer I/Os at the same recall.
At recall $0.95$ a batch costs \sys\ $901$, $1{,}849$, and $5{,}007$ I/Os.
\pipea\ issues $6.9{\times}$, $2.9{\times}$, and $2.8{\times}$ as many and GLRU $2.2{\times}$, $2.0{\times}$, and $1.8{\times}$ as many.

Fewer I/Os become throughput.
With many threads the SSD limits \sys.
On OpenResearcher at $128$ threads the device delivers $10.8$\,GB/s to \sys\ and $10.4$\,GB/s to \pipea.
\sys\ reads $0.57$\,MB per query and \pipea\ $3.9$\,MB, a $6.9{\times}$ gap close to the $7.1{\times}$ gap in QPS at $128$ threads.

Fewer I/Os explain little of the latency on one thread.
GLRU removes the same redundant I/Os as \sys.
At the same $L$ of $64$, $384$, and $256$ it issues $938$, $2{,}086$, and $5{,}550$ per batch against $856$, $1{,}953$, and $5{,}128$ for \sys, yet at these $L$ it is only $1.1$ to $1.2{\times}$ faster than \pipea.
One thread leaves the SSD far from its limit.
A \sys\ thread on OpenResearcher reads $0.29$\,GB/s, against $10.4$ to $10.8$\,GB/s at $128$ threads.
At these $L$ GLRU is as fast as \sys\ on one thread, at $57$, $26$, and $74$\,ms against $59$, $26$, and $73$\,ms.
The single-thread gain of \sys\ therefore comes from the smaller $L$ that exact-distance sharing allows (Figure~\ref{fig:variants}).
On one thread \sys\ serves $518$, $290$, and $424$ QPS at recall $0.95$, against $277$, $162$, and $260$ for \pipea.
Figure~\ref{fig:stage-profile} shows where its time goes.
PQ comparisons are the largest stage in every batch-size bucket at $51$ to $59\%$ of walk time, and they take $43\%$ and $47\%$ over all batches of QuAC+CANARD and NarrativeQA.
As batches grow, exact distances rise from $3\%$ to $25\%$ because each fetched node is scored for more queries, and I/O falls from $26\%$ to $7\%$.
Blocked waiting for I/Os is $1\%$, $11\%$, and $3\%$ of walk time on the three workloads with $4$ I/Os in flight per query, against $16\%$, $13\%$, and $21\%$ for \pipea\ with up to $32$ (Sec.~\ref{sec:bg:cache}).

\paragraph{Design choices.}
Figure~\ref{fig:variants} changes one design choice of \sys\ at a time.
Every variant runs at the $L$ where it reaches recall $0.95$.
The top row is the mean batch latency on one thread, and the bottom row is the QPS at the knee of the variant.
Each QPS is the median of three runs.
The first bar of every panel is \sys, and the label above every other bar is its value relative to \sys.

\paragraph{Exact-distance sharing.}
Without exact-distance sharing each query ranks only the nodes its own walk fetched.
Page sharing stays, so at equal $L$ the walk reads the same pages.
Recall at equal $L$ falls by $0.044$, $0.025$, and $0.029$, and recall $0.95$ needs $L$ of $184$, $688$, and $464$ instead of $70$, $304$, and $248$.
The larger $L$ raises latency by $1.6{\times}$, $1.9{\times}$, and $1.6{\times}$ and lowers throughput by $2.2{\times}$, $2.2{\times}$, and $1.9{\times}$.

\paragraph{Pipelined walk.}
\sys\ keeps up to four I/Os in flight for every query and processes a page as soon as it completes.
The hop-synchronous variant changes only this schedule.
It issues one round of I/Os for all queries and waits for the whole round before it processes any page.
With the same sharing it has $1.04{\times}$, $1.35{\times}$, and $1.08{\times}$ the latency and $1.07{\times}$, $1.08{\times}$, and $1.06{\times}$ lower throughput.

\paragraph{Sharing neighbors.}
\sys\ PQ-scores a fetched node's neighbors only for the queries that picked the node (Sec.~\ref{sec:design:noshare}).
Our first design scored them for every query in the batch, and the neighbor-sharing variant restores it.
At equal $L$ it computes $3.7{\times}$, $2.2{\times}$, and $3.2{\times}$ the PQ distances per query and reads $7$ to $10\%$ fewer pages, at a recall that is equal or up to $0.008$ higher.
On OpenResearcher this costs $2.4{\times}$ the latency on one thread and $2.9{\times}$ lower throughput, because PQ comparisons take more than half of a walk there (Figure~\ref{fig:stage-profile}).
On QuAC+CANARD and NarrativeQA recall $0.95$ needs $L$ of $289$ and $208$ instead of $304$ and $248$, so the variant reads $13\%$ and $23\%$ fewer pages per query.
It is still $1.3{\times}$ and $1.5{\times}$ slower on one thread, but its throughput at the knee is $1.19{\times}$ and $1.07{\times}$ that of \sys.
\sys\ keeps the selective design because it is faster on one thread on all three workloads and its throughput loss on OpenResearcher is larger than the gains on the other two.
We bound what a cheaper selection rule could gain.
We call a true top-$10$ neighbor of a query recoverable if it appears only in the neighbor lists of nodes that its batch-mates selected and its PQ distance would have entered the query's candidate list when such a node was read.
A rule that added exactly the recoverable neighbors at no cost would lower the single-thread latency and raise the throughput on $128$ threads at recall $0.90$ and $0.95$ by at most $1.1{\times}$, $1.2{\times}$, and $1.6{\times}$.
This first-order estimate comes from the hop-synchronous variant.
Selective sharing therefore has little room on OpenResearcher and QuAC+CANARD, and NarrativeQA leaves the most to a cheaper selection rule.

\paragraph{Sensitivity.}
Two parameters set how many I/Os and queries one thread interleaves, the I/Os in flight per query and the threads per batch $k$.
Every system runs at the $L$ where it reaches recall $0.95$.

\paragraph{I/Os in flight per query.}
The thread computes without waiting only if the I/Os in flight of the whole batch outnumber the pages it processes during the latency of one I/O.
A \pipea\ query on one thread has only its own I/Os, and raising them from $4$ to $32$ cuts its batch latency to $0.30$ to $0.54$ of the latency at $4$.
A \sys\ thread draws its I/Os from all $B$ walks of its batch, so $4$ per query give it up to $4B$ in flight, and this default is within $1.11{\times}$ of the lowest latency over $1$ to $16$ per query.
More I/Os per query add up to $20\%$ more pages, because a query selects all of its pending I/Os from the same candidate list before their neighbors arrive.
On $128$ threads the SSD is saturated, and the I/Os in flight matter less.

\paragraph{Threads per batch.}
Figure~\ref{fig:tladder} fixes each system's $L$ and adds threads step by step.
Each point is one thread count, placed at its QPS and the mean latency of a batch, and the black-edged point is the knee.
\sys\ has one line for each $k$ from $1$ to $16$.
A curve further to the lower right serves more queries at a lower batch latency.
A larger $k$ lowers the batch latency of \sys\ and also lowers its throughput.
With $k{=}16$ the batch latency at the knee is $15$, $12$, and $18$\,ms, $2.5{\times}$, $2.6{\times}$, and $1.6{\times}$ below \pipea.
The throughput falls to $89\%$, $46\%$, and $75\%$ of the $k{=}1$ knee, which is still $6.3{\times}$, $1.8{\times}$, and $2.6{\times}$ that of \pipea.
A thread that receives no query of its batch idles.
Every QuAC+CANARD batch has $4$ to $12$ queries, so with $k{=}16$ only $45\%$ of the threads hold a query on average.
The best $k$ depends on the batch sizes, the hardware, and the latency target.
On our server we recommend about four queries per thread of an average batch, the $k$ of Sec.~\ref{sec:eval:main}.

\paragraph{Memory.}
We measure the peak resident memory of each system at its knee, leaving out the query and ground-truth files of the benchmark.
A \sys\ thread keeps a record of every node its walks read until the batch ends, because a later pick of the same node needs the node's neighbor list.
With $k{=}1$ every query has its exact distance to a node once the node is processed, so the record holds only the neighbor list, $576$\,B on OpenResearcher and $320$\,B on the other two workloads (Sec.~\ref{sec:impl:records}).
The search then takes $38$, $6.6$, and $17$\,MiB per \sys\ thread, against $10$, $6.6$, and $10$\,MiB per \pipea\ thread.
With the index the process needs $5.0$, $3.0$, and $1.2$\,GiB, against $3.5$, $3.0$, and $0.8$\,GiB for \pipea\ and $5.2$, $3.5$, and $1.4$\,GiB for GLRU.
With the $k$ of Sec.~\ref{sec:eval:main} the other threads of the batch compute exact distances to a node after the walks end, so the record also keeps the node's vector, $16.6$\,KB on OpenResearcher and $3.3$\,KB on the other two.
A thread walks only its share of the batch, and \sys\ needs $4.6$, $3.3$, and $0.8$\,GiB.
Memory costs at most $1.4\%$ of the hourly price in Table~\ref{tab:cost}.

\subsection{More SSDs}
\label{sec:eval:scaling}

Our node has more cores than one SSD can feed, so on many threads both systems stop at the device (Figure~\ref{fig:tladder}).
We project what more SSDs would buy on OpenResearcher from two measured quantities per system, the peak QPS of one SSD and the QPS of one thread.
Throughput is the smaller of a device bound, which grows with the SSD count, and a CPU bound of $128$ threads times the per-thread QPS.
Each SSD carries $18.8$K QPS of \sys\ and $2.6$K of \pipea.
One \sys\ thread serves $510$ QPS on the batches of the thread ladder ($518$ on the first $1{,}000$ batches in Sec.~\ref{sec:eval:analysis}), and one \pipea\ thread with $32$ I/Os in flight serves $277$.
\sys\ runs out of cores between three and four SSDs, at about $65$K QPS.
\pipea\ needs about thirteen SSDs to keep the same threads busy and stops near $35$K QPS.
The ratio narrows from the peak ratio of $7.1{\times}$ on one SSD to the per-thread ratio of $1.8{\times}$ once both are bound by CPU.
The projection assumes that throughput grows linearly with threads, which \sys\ does within $7\%$ up to $16$ threads, and that a \pipea\ thread keeps its single-thread QPS.

%% file: sections/06_related_work.tex
\section{Related Work}
\label{sec:related}

\paragraph{Agentic retrieval workloads}
LLM agents turn one intent into many retrieval calls. Query rewriting and
HyDE issue $N$ paraphrases or hypothetical answers~\cite{queryrewrite,query2doc,hyde},
decomposition issues one call per sub-question~\cite{selfask,leasttomost,parallelsearch},
deep-research agents fan out across subagents~\cite{anthropic_multiagent,openai_deepresearch},
fact-checkers ground each generated sentence~\cite{factscore,safe}, and RL
post-training samples many rollouts per prompt~\cite{grpo,searchr1}. These
works treat retrieval as a black box and pay for each call in full.

\paragraph{Disk-resident graph ANN}
HNSW~\cite{hnsw} and NSG~\cite{nsg} dominate in-memory ANN
search~\cite{annsurvey}, and \diska~\cite{diskann} moves them to SSD by
paging a Vamana graph and pruning with in-RAM PQ codes. \pipea~\cite{pipeann} overlaps I/O with PQ,
Starling~\cite{starling} co-locates a walk's nodes on disk, and FreshDiskANN~\cite{freshdiskann} and
SPFresh~\cite{spfresh} add streaming updates. These target single-query
latency, so a batch reruns the search $B$ times, refetching the same pages,
and each query ranks only the nodes its own walk fetched. GPU indexes such as CAGRA~\cite{cagra} batch many
independent queries for throughput but still run each query's walk separately
rather than sharing it across a related batch. \sys\
keeps the index format of \diska\ and the pipelined I/Os of \pipea, reads each
page once per batch, and lets every query rank every node the batch fetched.

\paragraph{Multi-vector retrieval}
A parallel line of work indexes each item as a \emph{set} of token embeddings under the Chamfer (MaxSim) score~\cite{colbert}. 
Most systems pool every token into one single-vector index and query it per token, as in ColBERTv2~\cite{colbertv2}, PLAID~\cite{plaid}, and EMVB~\cite{emvb}.
MUVERA~\cite{muvera} instead encodes each set as one fixed-dimensional vector, DESSERT~\cite{dessert} as an LSH sketch, and IGP~\cite{igp} and GEM~\cite{gem} build proximity graphs over token sets. 
This line is orthogonal to \sys. 
Multi-vector retrieval returns one ranked list per set-valued query, whereas \sys\ serves a batch of single-vector point queries that share an intent but keep independent top-$K$ results. 

%% file: sections/07_discussion.tex
\section{Discussion and Limitations}
\label{sec:discussion}

\begin{figure}[t]
  \centering
  \includegraphics[width=\columnwidth]{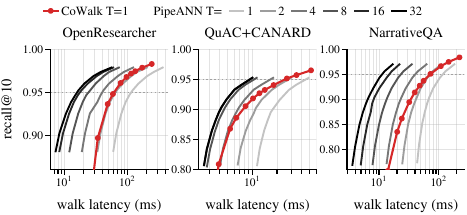}
  \caption{Batch-parallel \pipea\ on $T$ threads vs.\ single-thread \sys: recall@10 vs.\ walk latency.}
  \label{fig:latency-threads}
\end{figure}

\paragraph{Idle cores.}
A per-query system with idle cores can spread the $B$ queries of a batch over $T$ threads.
Figure~\ref{fig:latency-threads} compares this batch-parallel \pipea\ with \sys\ on one thread.
At recall $0.95$ batch-parallel \pipea\ on $2$ threads takes $65$, $27$, and $71$\,ms, against $62$, $25$, and $71$\,ms for single-thread \sys.
With $4$ threads it is faster on all three workloads, and with $32$ threads it finishes a batch in $16$, $9$, and $11$\,ms.
It pays for this with $T$ cores and unchanged I/Os per query, so it gives up the throughput of Sec.~\ref{sec:eval:main}.
\sys\ can also spread a batch over $k$ threads (Sec.~\ref{sec:eval:main}).
On $8$ threads with one batch at a time it finishes a batch in $10.8$, $10.2$, and $15.3$\,ms, against $24.3$, $10.5$, and $23.2$\,ms for batch-parallel \pipea\ on $8$ threads.

\begin{figure}[t]
  \centering
  \includegraphics[width=\columnwidth]{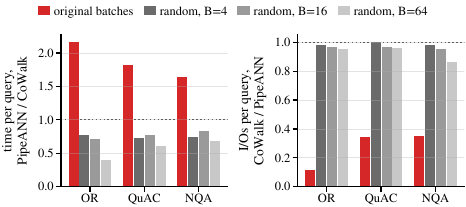}
  \caption{Original vs.\ random batches at recall $0.95$, single thread.}
  \label{fig:random-batches}
\end{figure}

\paragraph{Unrelated queries.}
\sys\ gains only when the queries of a batch search the same region of the graph.
The agentic batches of Sec.~\ref{sec:bg:similarity} have this property.
Batches of unrelated queries do not.
Figure~\ref{fig:random-batches} replaces each workload's batches with $B$ queries drawn at random from all its queries.
\sys\ then reads at most $14\%$ fewer pages per query than \pipea, against $65$ to $88\%$ fewer on the original batches, and its recall at equal $L$ is up to $0.015$ lower than that of \pipea.
On one thread \sys\ is then slower than \pipea.
At recall $0.95$ it spends $1.2$ to $1.4{\times}$ more time per query for $B$ of $4$ and $16$ and $1.5$ to $2.5{\times}$ more for $B{=}64$, while it spends $1.6$ to $2.2{\times}$ less on the original batches.
On $128$ threads, where the SSD limits both systems, its throughput on random batches of $16$ is $1.10{\times}$, $1.26{\times}$, and $1.28{\times}$ that of \pipea.
Unrelated batches thus make \sys\ slower than \pipea\ on one thread and remove most of its gain on many threads.

\paragraph{Read-path-only implementation.}
\sys\ shares the read path only. Concurrent inserts and deletes during a
walk fall back to the per-query update path of \pipea, and a joint
update path is left to future work. This fits our target setting, where
agent retrieval is read-dominated and index refresh runs in the background.

%% file: sections/08_conclusion.tex
\section{Conclusion}
\label{sec:conclusion}

LLM agents issue batches of related queries drawn from a single intent.
On three workloads, queries of the same batch lie $1.3$ to $2.4\times$ closer
than random pairs, and $31$ to $67\%$ of the I/Os of per-query search
fetch a page that a batch-mate already fetched. A page cache shared by the batch removes these I/Os but
leaves the walk of every query and its distance computations unchanged.
\sys\ runs the queries of a batch as cooperating walks on one thread. Every
query keeps its own candidate list, a page is read once per batch, and every
query ranks every node the batch fetched, so a query reaches its recall with a
shorter candidate list. At recall@10 of $0.95$, \sys\ answers a batch on one
thread $1.6$ to $1.9\times$ faster than the strongest per-query baseline, and
on a server it serves $3.0$ to $6.9\times$ its throughput at a similar batch
latency. We see \sys\ as a step toward retrieval back ends designed
for batched agent retrieval.

%% file: main.bbl

\begin{thebibliography}{38}

%%% the Web address.

\ifx \showCODEN    \undefined \def \showCODEN     #1{\unskip}     \fi
\ifx \showDOI      \undefined \def \showDOI       #1{#1}\fi
\ifx \showISBNx    \undefined \def \showISBNx     #1{\unskip}     \fi
\ifx \showISBNxiii \undefined \def \showISBNxiii  #1{\unskip}     \fi
\ifx \showISSN     \undefined \def \showISSN      #1{\unskip}     \fi
\ifx \showLCCN     \undefined \def \showLCCN      #1{\unskip}     \fi
\ifx \shownote     \undefined \def \shownote      #1{#1}          \fi
\ifx \showarticletitle \undefined \def \showarticletitle #1{#1}   \fi
\ifx \showURL      \undefined \def \showURL       {\relax}        \fi
% invisible to TeX
\providecommand\bibfield[2]{#2}
\providecommand\bibinfo[2]{#2}
\providecommand\natexlab[1]{#1}
\providecommand\showeprint[2][]{arXiv:#2}

\bibitem[{Anthropic}(2025a)]%
        {anthropic_skills}
\bibfield{author}{\bibinfo{person}{{Anthropic}}.}
  \bibinfo{year}{2025}\natexlab{a}.
\newblock \bibinfo{title}{Equipping Agents for the Real World with Agent
  Skills}.
\newblock
  \bibinfo{howpublished}{\url{https://www.anthropic.com/engineering/equipping-agents-for-the-real-world-with-agent-skills}}.
\newblock
\newblock
\shownote{Accessed 2026-06-09}.


\bibitem[{Anthropic}(2025b)]%
        {anthropic_multiagent}
\bibfield{author}{\bibinfo{person}{{Anthropic}}.}
  \bibinfo{year}{2025}\natexlab{b}.
\newblock \bibinfo{title}{How We Built Our Multi-Agent Research System}.
\newblock
  \bibinfo{howpublished}{\url{https://www.anthropic.com/engineering/multi-agent-research-system}}.
\newblock
\newblock
\shownote{Accessed 2026-09-22}.


\bibitem[Bian et~al\mbox{.}(2025)]%
        {igp}
\bibfield{author}{\bibinfo{person}{Zheng Bian}, \bibinfo{person}{Man~Lung Yiu},
  {and} \bibinfo{person}{Bo Tang}.} \bibinfo{year}{2025}\natexlab{}.
\newblock \showarticletitle{IGP: Efficient Multi-Vector Retrieval via Proximity
  Graph Index}. In \bibinfo{booktitle}{\emph{Proceedings of the 48th
  International ACM SIGIR Conference on Research and Development in Information
  Retrieval}} (Padua, Italy) \emph{(\bibinfo{series}{SIGIR '25})}.
  \bibinfo{publisher}{Association for Computing Machinery},
  \bibinfo{address}{New York, NY, USA}, \bibinfo{pages}{2524–2533}.
\newblock
\showISBNx{9798400715921}
\urldef\tempurl%
\url{https://doi.org/10.1145/3726302.3730004}
\showDOI{\tempurl}


\bibitem[Chen et~al\mbox{.}(2021)]%
        {spann}
\bibfield{author}{\bibinfo{person}{Qi Chen}, \bibinfo{person}{Bing Zhao},
  \bibinfo{person}{Haidong Wang}, \bibinfo{person}{Mingqin Li},
  \bibinfo{person}{Chuanjie Liu}, \bibinfo{person}{Zengzhong Li},
  \bibinfo{person}{Mao Yang}, {and} \bibinfo{person}{Jingdong Wang}.}
  \bibinfo{year}{2021}\natexlab{}.
\newblock \showarticletitle{SPANN: highly-efficient billion-scale approximate
  nearest neighbor search}. In \bibinfo{booktitle}{\emph{Proceedings of the
  35th International Conference on Neural Information Processing Systems}}
  \emph{(\bibinfo{series}{NIPS '21})}. \bibinfo{publisher}{Curran Associates
  Inc.}, \bibinfo{address}{Red Hook, NY, USA}, Article
  \bibinfo{articleno}{398}, \bibinfo{numpages}{14}~pages.
\newblock
\showISBNx{9781713845393}


\bibitem[Choi et~al\mbox{.}(2018)]%
        {quac}
\bibfield{author}{\bibinfo{person}{Eunsol Choi}, \bibinfo{person}{He He},
  \bibinfo{person}{Mohit Iyyer}, \bibinfo{person}{Mark Yatskar},
  \bibinfo{person}{Wen-tau Yih}, \bibinfo{person}{Yejin Choi},
  \bibinfo{person}{Percy Liang}, {and} \bibinfo{person}{Luke Zettlemoyer}.}
  \bibinfo{year}{2018}\natexlab{}.
\newblock \showarticletitle{{Q}u{AC}: Question Answering in Context}. In
  \bibinfo{booktitle}{\emph{Proceedings of the 2018 Conference on Empirical
  Methods in Natural Language Processing}},
  \bibfield{editor}{\bibinfo{person}{Ellen Riloff}, \bibinfo{person}{David
  Chiang}, \bibinfo{person}{Julia Hockenmaier}, {and}
  \bibinfo{person}{Jun{'}ichi Tsujii}} (Eds.). \bibinfo{publisher}{Association
  for Computational Linguistics}, \bibinfo{address}{Brussels, Belgium},
  \bibinfo{pages}{2174--2184}.
\newblock
\urldef\tempurl%
\url{https://doi.org/10.18653/v1/D18-1241}
\showDOI{\tempurl}


\bibitem[Dhulipala et~al\mbox{.}(2026)]%
        {muvera}
\bibfield{author}{\bibinfo{person}{Laxman Dhulipala}, \bibinfo{person}{Majid
  Hadian}, \bibinfo{person}{Rajesh Jayaram}, \bibinfo{person}{Jason Lee}, {and}
  \bibinfo{person}{Vahab Mirrokni}.} \bibinfo{year}{2026}\natexlab{}.
\newblock \bibinfo{title}{MUVERA: Multi-Vector Retrieval via Fixed Dimensional
  Encodings}.
\newblock
\newblock
\showeprint[arxiv]{2405.19504}~[cs.DS]
\urldef\tempurl%
\url{https://arxiv.org/abs/2405.19504}
\showURL{%
\tempurl}


\bibitem[Elgohary et~al\mbox{.}(2019)]%
        {canard}
\bibfield{author}{\bibinfo{person}{Ahmed Elgohary}, \bibinfo{person}{Denis
  Peskov}, {and} \bibinfo{person}{Jordan Boyd-Graber}.}
  \bibinfo{year}{2019}\natexlab{}.
\newblock \showarticletitle{Can You Unpack That? Learning to Rewrite
  Questions-in-Context}. In \bibinfo{booktitle}{\emph{Proceedings of the 2019
  Conference on Empirical Methods in Natural Language Processing and the 9th
  International Joint Conference on Natural Language Processing
  (EMNLP-IJCNLP)}}, \bibfield{editor}{\bibinfo{person}{Kentaro Inui},
  \bibinfo{person}{Jing Jiang}, \bibinfo{person}{Vincent Ng}, {and}
  \bibinfo{person}{Xiaojun Wan}} (Eds.). \bibinfo{publisher}{Association for
  Computational Linguistics}, \bibinfo{address}{Hong Kong, China},
  \bibinfo{pages}{5918--5924}.
\newblock
\urldef\tempurl%
\url{https://doi.org/10.18653/v1/D19-1605}
\showDOI{\tempurl}


\bibitem[Engels et~al\mbox{.}(2023)]%
        {dessert}
\bibfield{author}{\bibinfo{person}{Joshua Engels}, \bibinfo{person}{Benjamin
  Coleman}, \bibinfo{person}{Vihan Lakshman}, {and} \bibinfo{person}{Anshumali
  Shrivastava}.} \bibinfo{year}{2023}\natexlab{}.
\newblock \showarticletitle{{DESSERT}: An Efficient Algorithm for Vector Set
  Search with Vector Set Queries}. In \bibinfo{booktitle}{\emph{Thirty-seventh
  Conference on Neural Information Processing Systems}}.
\newblock
\urldef\tempurl%
\url{https://openreview.net/forum?id=kXfrlWXLwH}
\showURL{%
\tempurl}


\bibitem[Fu et~al\mbox{.}(2019)]%
        {nsg}
\bibfield{author}{\bibinfo{person}{Cong Fu}, \bibinfo{person}{Chao Xiang},
  \bibinfo{person}{Changxu Wang}, {and} \bibinfo{person}{Deng Cai}.}
  \bibinfo{year}{2019}\natexlab{}.
\newblock \showarticletitle{Fast approximate nearest neighbor search with the
  navigating spreading-out graph}.
\newblock \bibinfo{journal}{\emph{Proc. VLDB Endow.}} \bibinfo{volume}{12},
  \bibinfo{number}{5} (\bibinfo{date}{Jan.} \bibinfo{year}{2019}),
  \bibinfo{pages}{461–474}.
\newblock
\showISSN{2150-8097}
\urldef\tempurl%
\url{https://doi.org/10.14778/3303753.3303754}
\showDOI{\tempurl}


\bibitem[Gao et~al\mbox{.}(2023)]%
        {hyde}
\bibfield{author}{\bibinfo{person}{Luyu Gao}, \bibinfo{person}{Xueguang Ma},
  \bibinfo{person}{Jimmy Lin}, {and} \bibinfo{person}{Jamie Callan}.}
  \bibinfo{year}{2023}\natexlab{}.
\newblock \showarticletitle{Precise Zero-Shot Dense Retrieval without Relevance
  Labels}. In \bibinfo{booktitle}{\emph{Proceedings of the 61st Annual Meeting
  of the Association for Computational Linguistics (Volume 1: Long Papers)}},
  \bibfield{editor}{\bibinfo{person}{Anna Rogers}, \bibinfo{person}{Jordan
  Boyd-Graber}, {and} \bibinfo{person}{Naoaki Okazaki}} (Eds.).
  \bibinfo{publisher}{Association for Computational Linguistics},
  \bibinfo{address}{Toronto, Canada}, \bibinfo{pages}{1762--1777}.
\newblock
\urldef\tempurl%
\url{https://doi.org/10.18653/v1/2023.acl-long.99}
\showDOI{\tempurl}


\bibitem[Guo and Lu(2026)]%
        {pipeann}
\bibfield{author}{\bibinfo{person}{Hao Guo} {and} \bibinfo{person}{Youyou Lu}.}
  \bibinfo{year}{2026}\natexlab{}.
\newblock \showarticletitle{Achieving Low-Latency Graph-Based Vector Search via
  Aligning Best-First Search Algorithm with SSD}.
\newblock \bibinfo{journal}{\emph{ACM Trans. Storage}} \bibinfo{volume}{22},
  \bibinfo{number}{2}, Article \bibinfo{articleno}{12} (\bibinfo{date}{March}
  \bibinfo{year}{2026}), \bibinfo{numpages}{30}~pages.
\newblock
\showISSN{1553-3077}
\urldef\tempurl%
\url{https://doi.org/10.1145/3793926}
\showDOI{\tempurl}


\bibitem[Jin et~al\mbox{.}(2025)]%
        {searchr1}
\bibfield{author}{\bibinfo{person}{Bowen Jin}, \bibinfo{person}{Hansi Zeng},
  \bibinfo{person}{Zhenrui Yue}, \bibinfo{person}{Jinsung Yoon},
  \bibinfo{person}{Sercan Arik}, \bibinfo{person}{Dong Wang},
  \bibinfo{person}{Hamed Zamani}, {and} \bibinfo{person}{Jiawei Han}.}
  \bibinfo{year}{2025}\natexlab{}.
\newblock \bibinfo{title}{Search-R1: Training LLMs to Reason and Leverage
  Search Engines with Reinforcement Learning}.
\newblock
\newblock
\showeprint[arxiv]{2503.09516}~[cs.CL]
\urldef\tempurl%
\url{https://arxiv.org/abs/2503.09516}
\showURL{%
\tempurl}


\bibitem[Karpukhin et~al\mbox{.}(2020)]%
        {dpr}
\bibfield{author}{\bibinfo{person}{Vladimir Karpukhin}, \bibinfo{person}{Barlas
  Oguz}, \bibinfo{person}{Sewon Min}, \bibinfo{person}{Patrick Lewis},
  \bibinfo{person}{Ledell Wu}, \bibinfo{person}{Sergey Edunov},
  \bibinfo{person}{Danqi Chen}, {and} \bibinfo{person}{Wen-tau Yih}.}
  \bibinfo{year}{2020}\natexlab{}.
\newblock \showarticletitle{Dense Passage Retrieval for Open-Domain Question
  Answering}. In \bibinfo{booktitle}{\emph{Proceedings of the 2020 Conference
  on Empirical Methods in Natural Language Processing (EMNLP)}},
  \bibfield{editor}{\bibinfo{person}{Bonnie Webber}, \bibinfo{person}{Trevor
  Cohn}, \bibinfo{person}{Yulan He}, {and} \bibinfo{person}{Yang Liu}} (Eds.).
  \bibinfo{publisher}{Association for Computational Linguistics},
  \bibinfo{address}{Online}, \bibinfo{pages}{6769--6781}.
\newblock
\urldef\tempurl%
\url{https://doi.org/10.18653/v1/2020.emnlp-main.550}
\showDOI{\tempurl}


\bibitem[Khattab and Zaharia(2020)]%
        {colbert}
\bibfield{author}{\bibinfo{person}{Omar Khattab} {and} \bibinfo{person}{Matei
  Zaharia}.} \bibinfo{year}{2020}\natexlab{}.
\newblock \showarticletitle{ColBERT: Efficient and Effective Passage Search via
  Contextualized Late Interaction over BERT}. In
  \bibinfo{booktitle}{\emph{Proceedings of the 43rd International ACM SIGIR
  Conference on Research and Development in Information Retrieval}} (Virtual
  Event, China) \emph{(\bibinfo{series}{SIGIR '20})}.
  \bibinfo{publisher}{Association for Computing Machinery},
  \bibinfo{address}{New York, NY, USA}, \bibinfo{pages}{39–48}.
\newblock
\showISBNx{9781450380164}
\urldef\tempurl%
\url{https://doi.org/10.1145/3397271.3401075}
\showDOI{\tempurl}


\bibitem[Ko{\v{c}}isk{\'y} et~al\mbox{.}(2018)]%
        {narrativeqa}
\bibfield{author}{\bibinfo{person}{Tom{\'a}{\v{s}} Ko{\v{c}}isk{\'y}},
  \bibinfo{person}{Jonathan Schwarz}, \bibinfo{person}{Phil Blunsom},
  \bibinfo{person}{Chris Dyer}, \bibinfo{person}{Karl~Moritz Hermann},
  \bibinfo{person}{G{\'a}bor Melis}, {and} \bibinfo{person}{Edward
  Grefenstette}.} \bibinfo{year}{2018}\natexlab{}.
\newblock \showarticletitle{The {N}arrative{QA} Reading Comprehension
  Challenge}.
\newblock \bibinfo{journal}{\emph{Transactions of the Association for
  Computational Linguistics}}  \bibinfo{volume}{6} (\bibinfo{year}{2018}),
  \bibinfo{pages}{317--328}.
\newblock
\urldef\tempurl%
\url{https://doi.org/10.1162/tacl_a_00023}
\showDOI{\tempurl}


\bibitem[Li et~al\mbox{.}(2026)]%
        {openresearcher}
\bibfield{author}{\bibinfo{person}{Zhuofeng Li}, \bibinfo{person}{Dongfu
  Jiang}, \bibinfo{person}{Xueguang Ma}, \bibinfo{person}{Haoxiang Zhang},
  \bibinfo{person}{Ping Nie}, \bibinfo{person}{Yuyu Zhang},
  \bibinfo{person}{Kai Zou}, \bibinfo{person}{Jianwen Xie}, \bibinfo{person}{Yu
  Zhang}, {and} \bibinfo{person}{Wenhu Chen}.} \bibinfo{year}{2026}\natexlab{}.
\newblock \showarticletitle{{OpenResearcher: A Fully Open Pipeline for
  Long-Horizon Deep Research Trajectory Synthesis}}.
\newblock \bibinfo{journal}{\emph{arXiv preprint arXiv:2603.20278}}
  (\bibinfo{year}{2026}).
\newblock


\bibitem[Ma et~al\mbox{.}(2023)]%
        {queryrewrite}
\bibfield{author}{\bibinfo{person}{Xinbei Ma}, \bibinfo{person}{Yeyun Gong},
  \bibinfo{person}{Pengcheng He}, \bibinfo{person}{hai zhao}, {and}
  \bibinfo{person}{Nan Duan}.} \bibinfo{year}{2023}\natexlab{}.
\newblock \showarticletitle{Query Rewriting in Retrieval-Augmented Large
  Language Models}. In \bibinfo{booktitle}{\emph{The 2023 Conference on
  Empirical Methods in Natural Language Processing}}.
\newblock
\urldef\tempurl%
\url{https://openreview.net/forum?id=gXq1cwkUZc}
\showURL{%
\tempurl}


\bibitem[Malkov and Yashunin(2020)]%
        {hnsw}
\bibfield{author}{\bibinfo{person}{Yu~A. Malkov} {and} \bibinfo{person}{D.~A.
  Yashunin}.} \bibinfo{year}{2020}\natexlab{}.
\newblock \showarticletitle{Efficient and Robust Approximate Nearest Neighbor
  Search Using Hierarchical Navigable Small World Graphs}.
\newblock \bibinfo{journal}{\emph{IEEE Trans. Pattern Anal. Mach. Intell.}}
  \bibinfo{volume}{42}, \bibinfo{number}{4} (\bibinfo{date}{April}
  \bibinfo{year}{2020}), \bibinfo{pages}{824–836}.
\newblock
\showISSN{0162-8828}
\urldef\tempurl%
\url{https://doi.org/10.1109/TPAMI.2018.2889473}
\showDOI{\tempurl}


\bibitem[Min et~al\mbox{.}(2023)]%
        {factscore}
\bibfield{author}{\bibinfo{person}{Sewon Min}, \bibinfo{person}{Kalpesh
  Krishna}, \bibinfo{person}{Xinxi Lyu}, \bibinfo{person}{Mike Lewis},
  \bibinfo{person}{Wen-tau Yih}, \bibinfo{person}{Pang Koh},
  \bibinfo{person}{Mohit Iyyer}, \bibinfo{person}{Luke Zettlemoyer}, {and}
  \bibinfo{person}{Hannaneh Hajishirzi}.} \bibinfo{year}{2023}\natexlab{}.
\newblock \showarticletitle{{FA}ct{S}core: Fine-grained Atomic Evaluation of
  Factual Precision in Long Form Text Generation}. In
  \bibinfo{booktitle}{\emph{Proceedings of the 2023 Conference on Empirical
  Methods in Natural Language Processing}},
  \bibfield{editor}{\bibinfo{person}{Houda Bouamor}, \bibinfo{person}{Juan
  Pino}, {and} \bibinfo{person}{Kalika Bali}} (Eds.).
  \bibinfo{publisher}{Association for Computational Linguistics},
  \bibinfo{address}{Singapore}, \bibinfo{pages}{12076--12100}.
\newblock
\urldef\tempurl%
\url{https://doi.org/10.18653/v1/2023.emnlp-main.741}
\showDOI{\tempurl}


\bibitem[Mohoney et~al\mbox{.}(2023)]%
        {hqi}
\bibfield{author}{\bibinfo{person}{Jason Mohoney}, \bibinfo{person}{Anil
  Pacaci}, \bibinfo{person}{Shihabur~Rahman Chowdhury}, \bibinfo{person}{Ali
  Mousavi}, \bibinfo{person}{Ihab~F. Ilyas}, \bibinfo{person}{Umar~Farooq
  Minhas}, \bibinfo{person}{Jeffrey Pound}, {and} \bibinfo{person}{Theodoros
  Rekatsinas}.} \bibinfo{year}{2023}\natexlab{}.
\newblock \showarticletitle{High-Throughput Vector Similarity Search in
  Knowledge Graphs}.
\newblock \bibinfo{journal}{\emph{Proc. ACM Manag. Data}} \bibinfo{volume}{1},
  \bibinfo{number}{2}, Article \bibinfo{articleno}{197} (\bibinfo{date}{June}
  \bibinfo{year}{2023}), \bibinfo{numpages}{25}~pages.
\newblock
\urldef\tempurl%
\url{https://doi.org/10.1145/3589777}
\showDOI{\tempurl}


\bibitem[Nardini et~al\mbox{.}(2024)]%
        {emvb}
\bibfield{author}{\bibinfo{person}{Franco~Maria Nardini},
  \bibinfo{person}{Cosimo Rulli}, {and} \bibinfo{person}{Rossano Venturini}.}
  \bibinfo{year}{2024}\natexlab{}.
\newblock \showarticletitle{Efficient Multi-vector Dense Retrieval with Bit
  Vectors}. In \bibinfo{booktitle}{\emph{European Conference on Information
  Retrieval}}. Springer, \bibinfo{pages}{3--17}.
\newblock


\bibitem[Ootomo et~al\mbox{.}(2024)]%
        {cagra}
\bibfield{author}{\bibinfo{person}{Hiroyuki Ootomo}, \bibinfo{person}{Akira
  Naruse}, \bibinfo{person}{Corey Nolet}, \bibinfo{person}{Ray Wang},
  \bibinfo{person}{Tamas Feher}, {and} \bibinfo{person}{Yong Wang}.}
  \bibinfo{year}{2024}\natexlab{}.
\newblock \showarticletitle{{ CAGRA: Highly Parallel Graph Construction and
  Approximate Nearest Neighbor Search for GPUs }}. In
  \bibinfo{booktitle}{\emph{2024 IEEE 40th International Conference on Data
  Engineering (ICDE)}}. \bibinfo{publisher}{IEEE Computer Society},
  \bibinfo{address}{Los Alamitos, CA, USA}, \bibinfo{pages}{4236--4247}.
\newblock
\urldef\tempurl%
\url{https://doi.org/10.1109/ICDE60146.2024.00323}
\showDOI{\tempurl}


\bibitem[{OpenAI}(2025)]%
        {openai_deepresearch}
\bibfield{author}{\bibinfo{person}{{OpenAI}}.} \bibinfo{year}{2025}\natexlab{}.
\newblock \bibinfo{title}{Introducing Deep Research}.
\newblock
  \bibinfo{howpublished}{\url{https://openai.com/index/introducing-deep-research/}}.
\newblock
\newblock
\shownote{Accessed 2026-06-09}.


\bibitem[{Perplexity AI}(2025)]%
        {perplexity_deepresearch}
\bibfield{author}{\bibinfo{person}{{Perplexity AI}}.}
  \bibinfo{year}{2025}\natexlab{}.
\newblock \bibinfo{title}{Introducing Perplexity Deep Research}.
\newblock
  \bibinfo{howpublished}{\url{https://www.perplexity.ai/hub/blog/introducing-perplexity-deep-research}}.
\newblock
\newblock
\shownote{Accessed 2026-06-09}.


\bibitem[Press et~al\mbox{.}(2023)]%
        {selfask}
\bibfield{author}{\bibinfo{person}{Ofir Press}, \bibinfo{person}{Muru Zhang},
  \bibinfo{person}{Sewon Min}, \bibinfo{person}{Ludwig Schmidt},
  \bibinfo{person}{Noah~A. Smith}, {and} \bibinfo{person}{Mike Lewis}.}
  \bibinfo{year}{2023}\natexlab{}.
\newblock \showarticletitle{Measuring and Narrowing the Compositionality Gap in
  Language Models}. In \bibinfo{booktitle}{\emph{The 2023 Conference on
  Empirical Methods in Natural Language Processing}}.
\newblock
\urldef\tempurl%
\url{https://openreview.net/forum?id=feiAVaSXdb}
\showURL{%
\tempurl}


\bibitem[Santhanam et~al\mbox{.}(2022a)]%
        {plaid}
\bibfield{author}{\bibinfo{person}{Keshav Santhanam}, \bibinfo{person}{Omar
  Khattab}, \bibinfo{person}{Christopher Potts}, {and} \bibinfo{person}{Matei
  Zaharia}.} \bibinfo{year}{2022}\natexlab{a}.
\newblock \showarticletitle{PLAID: An Efficient Engine for Late Interaction
  Retrieval}. In \bibinfo{booktitle}{\emph{Proceedings of the 31st ACM
  International Conference on Information \& Knowledge Management}} (Atlanta,
  GA, USA) \emph{(\bibinfo{series}{CIKM '22})}. \bibinfo{publisher}{Association
  for Computing Machinery}, \bibinfo{address}{New York, NY, USA},
  \bibinfo{pages}{1747–1756}.
\newblock
\showISBNx{9781450392365}
\urldef\tempurl%
\url{https://doi.org/10.1145/3511808.3557325}
\showDOI{\tempurl}


\bibitem[Santhanam et~al\mbox{.}(2022b)]%
        {colbertv2}
\bibfield{author}{\bibinfo{person}{Keshav Santhanam}, \bibinfo{person}{Omar
  Khattab}, \bibinfo{person}{Jon Saad-Falcon}, \bibinfo{person}{Christopher
  Potts}, {and} \bibinfo{person}{Matei Zaharia}.}
  \bibinfo{year}{2022}\natexlab{b}.
\newblock \showarticletitle{{C}ol{BERT}v2: Effective and Efficient Retrieval
  via Lightweight Late Interaction}. In \bibinfo{booktitle}{\emph{Proceedings
  of the 2022 Conference of the North American Chapter of the Association for
  Computational Linguistics: Human Language Technologies}},
  \bibfield{editor}{\bibinfo{person}{Marine Carpuat},
  \bibinfo{person}{Marie-Catherine de~Marneffe}, {and}
  \bibinfo{person}{Ivan~Vladimir Meza~Ruiz}} (Eds.).
  \bibinfo{publisher}{Association for Computational Linguistics},
  \bibinfo{address}{Seattle, United States}, \bibinfo{pages}{3715--3734}.
\newblock
\urldef\tempurl%
\url{https://doi.org/10.18653/v1/2022.naacl-main.272}
\showDOI{\tempurl}


\bibitem[Shao et~al\mbox{.}(2024)]%
        {grpo}
\bibfield{author}{\bibinfo{person}{Zhihong Shao}, \bibinfo{person}{Peiyi Wang},
  \bibinfo{person}{Qihao Zhu}, \bibinfo{person}{Runxin Xu},
  \bibinfo{person}{Junxiao Song}, \bibinfo{person}{Xiao Bi},
  \bibinfo{person}{Haowei Zhang}, \bibinfo{person}{Mingchuan Zhang},
  \bibinfo{person}{Y.~K. Li}, \bibinfo{person}{Y. Wu}, {and}
  \bibinfo{person}{Daya Guo}.} \bibinfo{year}{2024}\natexlab{}.
\newblock \bibinfo{title}{DeepSeekMath: Pushing the Limits of Mathematical
  Reasoning in Open Language Models}.
\newblock
\newblock
\showeprint[arxiv]{2402.03300}~[cs.CL]
\urldef\tempurl%
\url{https://arxiv.org/abs/2402.03300}
\showURL{%
\tempurl}


\bibitem[Singh et~al\mbox{.}(2021)]%
        {freshdiskann}
\bibfield{author}{\bibinfo{person}{Aditi Singh}, \bibinfo{person}{Suhas~Jayaram
  Subramanya}, \bibinfo{person}{Ravishankar Krishnaswamy}, {and}
  \bibinfo{person}{Harsha~Vardhan Simhadri}.} \bibinfo{year}{2021}\natexlab{}.
\newblock \showarticletitle{FreshDiskANN: {A} Fast and Accurate Graph-Based
  {ANN} Index for Streaming Similarity Search}.
\newblock \bibinfo{journal}{\emph{CoRR}}  \bibinfo{volume}{abs/2105.09613}
  (\bibinfo{year}{2021}).
\newblock
\showeprint[arXiv]{2105.09613}
\urldef\tempurl%
\url{https://arxiv.org/abs/2105.09613}
\showURL{%
\tempurl}


\bibitem[Subramanya et~al\mbox{.}(2019)]%
        {diskann}
\bibfield{author}{\bibinfo{person}{Suhas~Jayaram Subramanya},
  \bibinfo{person}{Devvrit}, \bibinfo{person}{Rohan Kadekodi},
  \bibinfo{person}{Ravishankar Krishaswamy}, {and}
  \bibinfo{person}{Harsha~Vardhan Simhadri}.} \bibinfo{year}{2019}\natexlab{}.
\newblock \bibinfo{booktitle}{\emph{DiskANN: fast accurate billion-point
  nearest neighbor search on a single node}}.
\newblock \bibinfo{publisher}{Curran Associates Inc.}, \bibinfo{address}{Red
  Hook, NY, USA}.
\newblock


\bibitem[Tian et~al\mbox{.}(2026)]%
        {gem}
\bibfield{author}{\bibinfo{person}{Yao Tian}, \bibinfo{person}{Zhoujin Tian},
  \bibinfo{person}{Xi Zhao}, \bibinfo{person}{Ruiyuan Zhang}, {and}
  \bibinfo{person}{Xiaofang Zhou}.} \bibinfo{year}{2026}\natexlab{}.
\newblock \bibinfo{title}{GEM: A Native Graph-based Index for Multi-Vector
  Retrieval}.
\newblock
\newblock
\showeprint[arxiv]{2603.20336}~[cs.IR]
\urldef\tempurl%
\url{https://arxiv.org/abs/2603.20336}
\showURL{%
\tempurl}


\bibitem[Wang et~al\mbox{.}(2023)]%
        {query2doc}
\bibfield{author}{\bibinfo{person}{Liang Wang}, \bibinfo{person}{Nan Yang},
  {and} \bibinfo{person}{Furu Wei}.} \bibinfo{year}{2023}\natexlab{}.
\newblock \showarticletitle{Query2doc: Query Expansion with Large Language
  Models}. In \bibinfo{booktitle}{\emph{The 2023 Conference on Empirical
  Methods in Natural Language Processing}}.
\newblock
\urldef\tempurl%
\url{https://openreview.net/forum?id=QH4EMvwF8I}
\showURL{%
\tempurl}


\bibitem[Wang et~al\mbox{.}(2024)]%
        {starling}
\bibfield{author}{\bibinfo{person}{Mengzhao Wang}, \bibinfo{person}{Weizhi Xu},
  \bibinfo{person}{Xiaomeng Yi}, \bibinfo{person}{Songlin Wu},
  \bibinfo{person}{Zhangyang Peng}, \bibinfo{person}{Xiangyu Ke},
  \bibinfo{person}{Yunjun Gao}, \bibinfo{person}{Xiaoliang Xu},
  \bibinfo{person}{Rentong Guo}, {and} \bibinfo{person}{Charles Xie}.}
  \bibinfo{year}{2024}\natexlab{}.
\newblock \showarticletitle{Starling: An I/O-Efficient Disk-Resident Graph
  Index Framework for High-Dimensional Vector Similarity Search on Data
  Segment}.
\newblock \bibinfo{journal}{\emph{Proc. ACM Manag. Data}} \bibinfo{volume}{2},
  \bibinfo{number}{1}, Article \bibinfo{articleno}{14} (\bibinfo{date}{March}
  \bibinfo{year}{2024}), \bibinfo{numpages}{27}~pages.
\newblock
\urldef\tempurl%
\url{https://doi.org/10.1145/3639269}
\showDOI{\tempurl}


\bibitem[Wang et~al\mbox{.}(2021)]%
        {annsurvey}
\bibfield{author}{\bibinfo{person}{Mengzhao Wang}, \bibinfo{person}{Xiaoliang
  Xu}, \bibinfo{person}{Qiang Yue}, {and} \bibinfo{person}{Yuxiang Wang}.}
  \bibinfo{year}{2021}\natexlab{}.
\newblock \showarticletitle{A comprehensive survey and experimental comparison
  of graph-based approximate nearest neighbor search}.
\newblock \bibinfo{journal}{\emph{Proc. VLDB Endow.}} \bibinfo{volume}{14},
  \bibinfo{number}{11} (\bibinfo{date}{July} \bibinfo{year}{2021}),
  \bibinfo{pages}{1964–1978}.
\newblock
\showISSN{2150-8097}
\urldef\tempurl%
\url{https://doi.org/10.14778/3476249.3476255}
\showDOI{\tempurl}


\bibitem[Wei et~al\mbox{.}(2024)]%
        {safe}
\bibfield{author}{\bibinfo{person}{Jerry Wei}, \bibinfo{person}{Chengrun Yang},
  \bibinfo{person}{Xinying Song}, \bibinfo{person}{Yifeng Lu},
  \bibinfo{person}{Nathan~Zixia Hu}, \bibinfo{person}{Jie Huang},
  \bibinfo{person}{Dustin Tran}, \bibinfo{person}{Daiyi Peng},
  \bibinfo{person}{Ruibo Liu}, \bibinfo{person}{Da Huang},
  \bibinfo{person}{Cosmo Du}, {and} \bibinfo{person}{Quoc~V Le}.}
  \bibinfo{year}{2024}\natexlab{}.
\newblock \showarticletitle{Long-form factuality in large language models}. In
  \bibinfo{booktitle}{\emph{The Thirty-eighth Annual Conference on Neural
  Information Processing Systems}}.
\newblock
\urldef\tempurl%
\url{https://openreview.net/forum?id=4M9f8VMt2C}
\showURL{%
\tempurl}


\bibitem[Xu et~al\mbox{.}(2023)]%
        {spfresh}
\bibfield{author}{\bibinfo{person}{Yuming Xu}, \bibinfo{person}{Hengyu Liang},
  \bibinfo{person}{Jin Li}, \bibinfo{person}{Shuotao Xu}, \bibinfo{person}{Qi
  Chen}, \bibinfo{person}{Qianxi Zhang}, \bibinfo{person}{Cheng Li},
  \bibinfo{person}{Ziyue Yang}, \bibinfo{person}{Fan Yang},
  \bibinfo{person}{Yuqing Yang}, \bibinfo{person}{Peng Cheng}, {and}
  \bibinfo{person}{Mao Yang}.} \bibinfo{year}{2023}\natexlab{}.
\newblock \showarticletitle{SPFresh: Incremental In-Place Update for
  Billion-Scale Vector Search}. In \bibinfo{booktitle}{\emph{Proceedings of the
  29th Symposium on Operating Systems Principles}} (Koblenz, Germany)
  \emph{(\bibinfo{series}{SOSP '23})}. \bibinfo{publisher}{Association for
  Computing Machinery}, \bibinfo{address}{New York, NY, USA},
  \bibinfo{pages}{545–561}.
\newblock
\showISBNx{9798400702297}
\urldef\tempurl%
\url{https://doi.org/10.1145/3600006.3613166}
\showDOI{\tempurl}


\bibitem[Zhao et~al\mbox{.}(2025)]%
        {parallelsearch}
\bibfield{author}{\bibinfo{person}{Shu Zhao}, \bibinfo{person}{Tan Yu},
  \bibinfo{person}{Anbang Xu}, \bibinfo{person}{Japinder Singh},
  \bibinfo{person}{Aaditya Shukla}, {and} \bibinfo{person}{Rama Akkiraju}.}
  \bibinfo{year}{2025}\natexlab{}.
\newblock \bibinfo{title}{ParallelSearch: Train your LLMs to Decompose Query
  and Search Sub-queries in Parallel with Reinforcement Learning}.
\newblock
\newblock
\showeprint[arxiv]{2508.09303}~[cs.CL]
\urldef\tempurl%
\url{https://arxiv.org/abs/2508.09303}
\showURL{%
\tempurl}


\bibitem[Zhou et~al\mbox{.}(2023)]%
        {leasttomost}
\bibfield{author}{\bibinfo{person}{Denny Zhou}, \bibinfo{person}{Nathanael
  Sch{\"a}rli}, \bibinfo{person}{Le Hou}, \bibinfo{person}{Jason Wei},
  \bibinfo{person}{Nathan Scales}, \bibinfo{person}{Xuezhi Wang},
  \bibinfo{person}{Dale Schuurmans}, \bibinfo{person}{Claire Cui},
  \bibinfo{person}{Olivier Bousquet}, \bibinfo{person}{Quoc~V Le}, {and}
  \bibinfo{person}{Ed~H. Chi}.} \bibinfo{year}{2023}\natexlab{}.
\newblock \showarticletitle{Least-to-Most Prompting Enables Complex Reasoning
  in Large Language Models}. In \bibinfo{booktitle}{\emph{The Eleventh
  International Conference on Learning Representations}}.
\newblock
\urldef\tempurl%
\url{https://openreview.net/forum?id=WZH7099tgfM}
\showURL{%
\tempurl}


\end{thebibliography}
